\documentclass[twocolumn, jcp, superscriptaddress]{revtex4}

\usepackage{mathtools}
\usepackage{algpseudocode}
\usepackage{algorithm}
\usepackage[caption=false]{subfig}
\usepackage{graphicx}
\begin{document}

\title{Linear-scaling time-dependent density-functional theory beyond the Tamm-Dancoff approximation: obtaining efficiency and accuracy with \emph{in situ} optimised local orbitals}
\author{T. J. Zuehlsdorff}
\email{tjz21@cam.ac.uk}
\affiliation{Cavendish Laboratory, J. J. Thomson Avenue, Cambridge CB3 0HE, UK}
\author{N. D. M. Hine}
\affiliation{Department of Physics, University of Warwick, Coventry, CV4 7AL}
\author{M. C. Payne}
\affiliation{Cavendish Laboratory, J. J. Thomson Avenue, Cambridge CB3 0HE, UK}
\author{P. D. Haynes}
\affiliation{Department of Materials, Imperial College London, Exhibition Road, London SW7 2AZ, UK}
\affiliation{Department of Physics, Imperial College London, Exhibition Road, London SW7 2AZ, UK}
\affiliation{Thomas Young Centre for Theory and Simulation of Materials, Imperial College London, Exhibition Road, London SW7 2AZ, UK}

\begin{abstract}
We present a solution of the full time-dependent density-functional theory (TDDFT) eigenvalue equation in the linear response formalism exhibiting a linear-scaling computational complexity with system size, without relying on the simplifying Tamm-Dancoff approximation (TDA). The implementation relies on representing the occupied and unoccupied subspace with two different sets of \emph{in situ} optimised localised functions, yielding a very compact and efficient representation of the transition density matrix of the excitation with the accuracy associated with a systematic basis set. The TDDFT eigenvalue equation is solved using a preconditioned conjugate gradient algorithm that is very memory-efficient. The algorithm is validated on a small test molecule and a good agreement with results obtained from standard quantum chemistry packages is found, with the preconditioner yielding a significant improvement in convergence rates. The method developed in this work is then used to reproduce experimental results of the absorption spectrum of bacteriochlorophyll in an organic solvent, where it is demonstrated that the TDA fails to reproduce the main features of the low energy spectrum, while the full TDDFT equation yields results in good qualitative agreement with experimental data. Furthermore, the need for explicitly including parts of the solvent into the TDDFT calculations is highlighted, making the treatment of large system sizes necessary that are well within reach of the capabilities of the algorithm introduced here. Finally, the linear-scaling properties of the algorithm are demonstrated by computing the lowest excitation energy of bacteriochlorophyll in solution. The largest systems considered in this work are of the same order of magnitude as a variety of widely studied pigment-protein complexes, opening up the possibility of studying their properties without having to resort to any semiclassical approximations to parts of the protein environment.
\end{abstract}

\date{\today}

\maketitle

\section{Introduction}
The study of optical properties of large complex systems is of increasing interest in computational biology, with most efforts being focused on understanding large pigment-protein complexes (PPCs)\cite{shim_fmo, neugebauer_fmo,danny_fmo, yellow_protein,neugebauer_summary}. These systems turn up in a variety of different roles in nature, from biosensors to light-harvesting and linker complexes in photosynthetic bacteria and \emph{ab-initio} computational studies can play a key role in gaining a deeper insight into the mechanisms governing them. However, PPCs are generally characterised by the fact that the protein environment plays an important role in influencing the absorption properties of the pigment, creating the need for large-scale quantum mechanical calculations that are computationally challenging\cite{neugebauer_fmo,danny_fmo,yellow_protein}. In general, time-dependent density-functional theory (TDDFT)\cite{runge_gross}, the time-dependent extension to ground-state density-functional theory (DFT)\cite{hohenberg_kohn,kohn_sham}, is considered the method of choice when treating this class of systems, mainly due to the good balance between computational cost and achievable accuracy for most common choices of exchange-correlation functionals.

In recent years there have been a number of developments in computational algorithms\cite{turbo,sternheimer_tddft,giustino} that have helped to make medium-sized systems routinely accessible to TDDFT. However, most common approaches to solving the low energy spectrum of a system using TDDFT show a computational complexity of at least $\mathcal{O}(N^3)$ with system size, imposing an upper limit on the system sizes that can be realistically studied and effectively ruling out a full treatment of the PPCs mentioned above. To treat these large biological systems explicitly in TDDFT, it is necessary to make use of computational approaches that scale linearly with system size. 

TDDFT is generally considered in two different flavours. The time domain approach, where the Kohn-Sham equations are propagated explicitly in time\cite{real_time}, and the linear response approach\cite{lr_tddft}, where the excitation energies of the system can be recast as the solutions to an effective eigenvalue equation\cite{casida_tddft, onida}. The time-domain approach can yield the entire spectrum of the system via a Fourier transform to the frequency domain, is non-perturbative and can thus be applied to problems beyond the linear response regime. However, it comes with the disadvantage that the Kohn-Sham equations have to be propagated sufficiently long to obtain narrow line-widths and dark states cannot be resolved. The direct solution of the TDDFT eigenvalue problem on the other hand can be used to obtain dark states and triplet transitions that are of interest in some photochemical processes. For this reason, we focus on the linear response flavour of TDDFT for the purpose of this work. In the time-domain TDDFT approach, an $\mathcal{O}(N)$ computational effort with system size can be achieved by extending linear-scaling techniques known from ground-state DFT to the time-dependent Kohn-Sham equations\cite{ls_real_time,ls_real_time2,bowler}. In the linear response approach, algorithms capable of solving for the lowest eigenvalues of the TDDFT eigenvalue equation are also known\cite{tretiak_tddft,tddft_onetep,challacombe_tddft}, opening up the possibility of a direct computation of excited states in large pigment-protein complexes without relying on additional semi-classical approximations. 

The linear response TDDFT equation is a non-Hermitian eigenvalue problem, causing it to be difficult to solve using standard off-the-shelf eigenvalue solvers. A simplifying approximation, known as the Tamm-Dancoff approximation (TDA)\cite{TDA,TDA_head_gordon}, recasts the problem into an Hermitian one but its effect on excitation energies and oscillator strengths is not straightforwardly understood. In this work, we introduce a linear-scaling implementation of full TDDFT in the framework of the $\mathtt{ONETEP}$ code\cite{onetep}, without relying on the TDA, as was required in a previous approach\cite{tddft_onetep}. We test the performance of the algorithm on a number of large scale systems and specifically investigate the quality of the full TDDFT eigenstates to those obtained within the TDA. The largest systems considered explicitly in this work are of similar size as a number of widely studied PPCs (see for example \cite{neugebauer_fmo,danny_fmo}), thus highlighting the capabilities of the algorithm developed here to enable a fully \emph{ab-initio} treatment of this class of systems. 

This work is organised as follows: Section~\ref{sec:background} focuses on providing a short overview of the theoretical background necessary for the main results presented in this work, with sections \ref{sec:lr_tddft} and \ref{sec:tsiper} introducing the linear response formalism to TDDFT, both in the form of a matrix eigenvalue equation and an effective variational principle. Section~\ref{sec:TDA} then provides a short summary of the $\mathtt{ONETEP}$ code in which the algorithm presented is implemented, as well as an overview over a solution to the Tamm-Dancoff eigenvalue problem\cite{tddft_onetep}. In section \ref{sec:full_tddft}, the linear-scaling solution to the full TDDFT eigenvalue equation is outlined, with a special focus being placed on the appropriate choice of preconditioner (\ref{sec:precond}) for the conjugate gradient algorithm. The power of the methodology developed here is demonstrated on a small test system by comparison to accurate benchmark results (\ref{sec:azo}), before moving on to a realistic system of bacteriochlorophyll $a$ in an organic solvent. It is shown that the algorithm developed here scales fully linearly with system size and allows the treatment of systems inaccessible by conventional approaches. 

\section{Theoretical background}
\label{sec:background}
In this section we briefly introduce the theoretical background of linear response TDDFT. We consider a Kohn-Sham system with ground-state density $\rho^{\{0\}}$ and occupied and unoccupied Kohn-Sham states $\{\psi_{v\sigma}^{\textrm{\scriptsize{KS}}}\}$ and $\{\psi_{c\sigma}^{\textrm{\scriptsize{KS}}}\}$ respectively, where $\sigma$ denotes a spin index. We limit the discussion to isolated systems, such that the Kohn-Sham eigenstates can be chosen to be real. Furthermore, only semi-local exchange-correlation functionals in the adiabatic approximation will be considered, thus ignoring any long-range and memory effects. While memory effects are routinely ignored in standard TDDFT implementations, long-range interactions can be included in form of hybrid functionals and are known to yield a better description for excitations in infinite systems and charge-transfer states, where semi-local exchange correlation functionals are known to fail\cite{onida,hybrid_functional}. However, for the purpose of this work the main focus is on excitations that retain a localised character and that are thus well described by semi-local functionals. 

\subsection{Linear response TDDFT}
\label{sec:lr_tddft}
In linear response TDDFT, the individual excitation energies of the system can be obtained by solving a non-Hermitian eigenvalue equation of the form\cite{casida_tddft,onida}
\begin{equation}
\begin{pmatrix} \textbf{A} & \textbf{B}\\ -\textbf{B} & -\textbf{A} \end{pmatrix}  \begin{pmatrix} \textbf{X} \\ \textbf{Y}  \end{pmatrix}= \omega \begin{pmatrix} \textbf{X} \\ \textbf{Y}  \end{pmatrix}
\label{eqn:TDDFT_ham}
\end{equation}
Here, $\textbf{A}$ and $\textbf{B}$ denote block matrices that can be conveniently expanded in a basis of unoccupied and occupied eigenstates of the ground state Kohn-Sham system:
\begin{eqnarray}
A_{cv\sigma,c'v'\sigma '}&=&\delta_{\sigma\sigma '}\delta_{cc'}\delta_{vv'}(\epsilon^{\textrm{\scriptsize{KS}}}_{c'\sigma '}-\epsilon^{\textrm{\scriptsize{KS}}}_{v'\sigma '})+K_{cv \sigma,c'v' \sigma '} \\
B_{cv\sigma,c'v'\sigma '}&=&K_{cv\sigma,c'v' \sigma '}.
\end{eqnarray}
where $\{\epsilon_{c\sigma}^{\textrm{\scriptsize{KS}}}\}$ and $\{\epsilon_{v\sigma}^{\textrm{\scriptsize{KS}}}\}$ denote the eigenvalues associated with the unoccupied and occupied Kohn-Sham states respectively. The eigenvectors are made up of two different components $\textbf{X}$ and $\textbf{Y}$ that can be thought of as excitation and de-exitation contributions to the corresponding eigenstates. As can be seen from Eq.~\ref{eqn:TDDFT_ham}, the effective non-Hermitian TDDFT eigenvalue matrix can be characterised by a diagonal part consisting of Kohn-Sham transitions between occupied and unoccupied states and off-diagonal coupling terms described through the coupling matrix $\textbf{K}$. The exact form of $\textbf{K}$ depends upon the exchange-correlation functional used. Here, we will limit our attention to (semi)-local functionals in the adiabatic approximation, in which case the matrix elements of $\textbf{K}$ can be expressed as
\begin{eqnarray} \nonumber
K_{cv\sigma,c'v'\sigma'}=\int \mathrm{d}^3 r\,\mathrm{d}^3r'\, \psi^{\textrm{\scriptsize{KS}}}_{c\sigma} (\textbf{r})\psi^{\textrm{\scriptsize{KS}}}_{v\sigma} (\textbf{r}) \\ \times\left[\frac{1}{|\textbf{r}-\textbf{r}'|}+\left.\frac{\delta^2 E_{\textrm{\scriptsize{xc}}}}{\delta \rho_\sigma(\textbf{r})\delta \rho_{\sigma'}(\textbf{r}')}\right|_{\rho^{\{0\}}}\right]\psi^{\textrm{\scriptsize{KS}}}_{c'\sigma '} (\textbf{r}')\psi^{\textrm{\scriptsize{KS}}}_{v'\sigma '} (\textbf{r}')
\end{eqnarray}
where $E_{\textrm{\scriptsize{xc}}}$ is the exchange-correlation energy. In the remaining part of this work, all spin indices will be dropped for convenience. 

From the structure of  Eq.~\ref{eqn:TDDFT_ham} it can be seen that the TDDFT eigenvalue equation has solutions in the positive and negative frequency domain. These positive and negative frequencies can be interpreted as excitation and de-excitation energies \cite{onida}. Note that due to the structure of the equation, the positive and negative eigenvalue solutions are coupled via the block matrix $\textbf{B}$. Assuming that the coupling of excitations to the de-excitation part of the full TDDFT spectrum is small, one can set the coupling matrices $\textbf{B}$ to zero, which causes a complete decoupling of the excitation and de-excitation part of the spectrum. The positive excitation energies can then be solved for via the Hermitian eigenvalue equation
\begin{equation}
\textbf{A}\textbf{X}=\omega \textbf{X}.
\label{eqn:TDA}
\end{equation}
Solving Eq.~\ref{eqn:TDA} instead of Eq.~\ref{eqn:TDDFT_ham} is referred to as the Tamm-Dancoff approximation (TDA)\cite{TDA,TDA_head_gordon}. While the TDA is often reported to yield reliable excitation energies in many situations\cite{TDA_head_gordon}, the structure of the equation violates both time-reversal symmetry and important sum rules related to the oscillator strengths\cite{thomas_reiche_kuhne} of the excitations. Furthermore, there are known cases where the TDA yields to significant errors in excitation energies\cite{azobenzene_benchmark}, making a treatment of the full eigenvalue problem desirable. However, as will be discussed in more detail in the next section, the main disadvantage of a full treatment of the TDDFT eigenvalue equation originates from the fact that Eq.~\ref{eqn:TDDFT_ham} constitutes a non-Hermitian eigenvalue problem, meaning that a variety of standard numerical methods for computing the eigenvalues of large matrices cannot be straightforwardly applied to it. 

\subsection{Iterative solutions to the TDDFT equation}
\label{sec:tsiper}

The dimensions of the TDDFT eigenvalue equation grow as $\mathcal{O}(N^2)$ with system size, making a direct diagonalisation of the matrices in Eq.~\ref{eqn:TDDFT_ham} or Eq.~\ref{eqn:TDA} undesirable for larger systems. Furthermore, one is often interested in a relatively small number of low energy excited states in the visible or ultraviolet energy range of a system, such that the computation of high energy excited states is unnecessary. The best approach to tackle the TDDFT eigenvalue problem for real systems of interest is thus to use iterative methods in order to calculate the lowest few excited states. 

Within the TDA, such an iterative scheme is straightforwardly defined, since the Hermitian properties of the block matrix $\textbf{A}$ allow for the definition of the lowest excitation of the system via a variational principle:
\begin{equation}
\omega^{\textrm{\scriptsize{TDA}}}_{\textrm{\scriptsize{min}}}=\min_{\textbf{X}}\Omega_{\textrm{\scriptsize{TDA}}}(\textbf{X})=\min_{\textbf{X}}\frac{\textbf{X}^\dagger \textbf{A}\textbf{X}}{\textbf{X}^\dagger\textbf{X}}.
\label{eqn:variational_TDA}
\end{equation}
From this definition, the gradient of the Rayleigh-Ritz functional $\Omega_{\textrm{\scriptsize{TDA}}}(\textbf{X})$ with respect to $\textbf{X}$ can then be straightforwardly computed
\begin{equation}
\frac{\partial \Omega_{\textrm{\scriptsize{TDA}}}(\textbf{X})}{\partial \textbf{X}}=\textbf{g}_{\textrm{\scriptsize{TDA}}}=\frac{2}{\textbf{X}^\dagger\textbf{X}}\left[\textbf{A}\textbf{X}- \frac{\textbf{X}^\dagger \textbf{A}\textbf{X}}{\textbf{X}^\dagger\textbf{X}} \textbf{X}\right].
\label{eqn:grad_TDA}
\end{equation}
This gradient can be used as a steepest-descent search direction in a conjugate gradient algorithm to optimise a random starting vector $\textbf{X}_{\textrm{\scriptsize{guess}}}$ until the lowest eigenstate $\omega^{\textrm{\scriptsize{TDA}}}_\textrm{\scriptsize{min}}$ has been obtained. Note that in order to compute the gradient in Eq.~\ref{eqn:grad_TDA}, it is only required to evaluate the matrix-vector product $\textbf{A}\textbf{X}$. This means that $\textbf{A}$ does not have to be explicitly calculated or stored during the calculation, which is prohibitive for large systems. 

The definition of the lowest eigenvalue of the Tamm-Dancoff TDDFT equation via the variational principle relies on the fact that it constitutes an Hermitian eigenvalue problem and is thus not generally possible for the full TDDFT equation. It was however pointed out by Thouless\cite{thouless} that since the blocks $\textbf{A}$ and $\textbf{B}$ of the full TDDFT equation are Hermitian, a variational principle can be formulated for its lowest \emph{positive} eigenstate via:
\begin{eqnarray} \nonumber
\omega_{\textrm{\scriptsize{min}}}&=&\min_{\left(\textbf{X} \, \textbf{Y} \right)} \Omega_{\textrm{\scriptsize{Thou}}}(\textbf{X},\textbf{Y}) \\
&=&\min_{\left(\textbf{X}\, \textbf{Y} \right)}\frac{\begin{pmatrix}\textbf{X}^\dagger & \textbf{Y}^\dagger \end{pmatrix}\begin{pmatrix} \textbf{A} & \textbf{B} \\ \textbf{B} & \textbf{A} \end{pmatrix} \begin{pmatrix}\textbf{X} \\ \textbf{Y} \end{pmatrix}}{|(\textbf{X}^\dagger\textbf{X})-(\textbf{Y}^\dagger\textbf{Y})|}.
\end{eqnarray}
The numerator of the above expression is guaranteed to be positive semi-definite for (semi)-local exchange-functionals, while the denominator is forced to be positive by taking the absolute value. 

While it is possible to use the above variational principle to directly obtain the lowest excited state of the full TDDFT equation, the Thouless functional is not the most ideal formulation of the problem in the context of semi-local exchange-correlation functionals (see Sec.~\ref{sec:operator}). A more computationally efficient reformulation was introduced by Tsiper \cite{tsiper} by noticing the equivalence of the full TDDFT eigenvalue problem to that of a set of classical harmonic oscillators. As a first step, one introduces the effective vectors $\textbf{p}=\textbf{X}-\textbf{Y}$ and $\textbf{q}=\textbf{X}+\textbf{Y}$. It can then be easily shown that the Thouless functional can be rewritten as:
\begin{eqnarray} \nonumber
\omega_{\textrm{\scriptsize{min}}}&=&\min_{\left(\textbf{p} \, \textbf{q} \right)} \Omega_{\textrm{\scriptsize{Tsip}}}(\textbf{p},\textbf{q}) \\
&=&\min_{\left(\textbf{p} \, \textbf{q} \right)}\frac{\begin{pmatrix}\textbf{p}^\dagger & \textbf{q}^\dagger \end{pmatrix}\begin{pmatrix} \textbf{A}-\textbf{B} & \textbf{0} \\ \textbf{0} & \textbf{A}+\textbf{B} \end{pmatrix} \begin{pmatrix}\textbf{p} \\ \textbf{q} \end{pmatrix}}{\left| \textbf{p}^\dagger \textbf{q} + \textbf{q}^\dagger \textbf{p} \right |}.
\label{eqn:tsiper_func}
\end{eqnarray}
Here, the variational principle is again only defined for the lowest \emph{positive} excitation of the system. Just like in the Thouless functional, the numerator of the functional is guaranteed to be positive-semidefinite. The denominator has to be forced to stay positive by taking the absolute value, since $\textbf{p}^\dagger \textbf{q} + \textbf{q}^\dagger \textbf{p}$  is not guaranteed to be positive-semidefinite. 

While the variational principles in Eqns. \ref{eqn:variational_TDA} and \ref{eqn:tsiper_func} are written in terms of the lowest excitation of the system only, the concept can be straightforwardly extended to higher excitations. In order to converge the second-lowest excitation $\omega_2$ of the system, the same effective functional of Eqn. \ref{eqn:tsiper_func} can be used in full TDDFT, with the additional constraint that $\textbf{p}_2$ and $\textbf{q}_2$, the trial vectors associated with $\omega_2$ obey an effective orthogonality constraint of the form
\begin{equation}
\left| \textbf{p}_1^\dagger \textbf{q}_2 + \textbf{q}_1^\dagger \textbf{p}_2 \right |=0
\label{eqn:tsiper_ortho}
\end{equation}
where $\textbf{p}_1$ and $\textbf{q}_1$ are taken to denote the vectors associated with $\omega_{\textrm{\scriptsize{min}}}$, the lowest excitation of the system. The principle can be extended to an arbitrary number of excited states to be converged, where the vectors $\textbf{p}_i$  and $\textbf{q}_i$ associated with the $i\textsuperscript{th}$ excitation are constrained to be orthogonal to the vectors of all other excitations via Eqn. \ref{eqn:tsiper_ortho}. 

\subsection{Tamm-Dancoff TDDFT in the $\mathtt{ONETEP}$ code}
\label{sec:TDA}
The linear-scaling solution to the full TDDFT equation developed in this work (see Sec.~\ref{sec:full_tddft}) is implemented in the $\mathtt{ONETEP}$ code\cite{onetep}. As with other linear-scaling DFT approaches, any reference to individual Kohn-Sham states $\{\psi_i^{\textrm{\scriptsize{KS}}}\}$ is given up in favour of a collective representation in the form of the single particle density matrix $\rho^{\{v\}}(\textbf{r},\textbf{r}')$ such that
\begin{eqnarray} \nonumber
\rho^{\{v\}}(\textbf{r},\textbf{r}')&=&\sum^{N_{\textrm{\scriptsize{occ}}}}_{v}\psi^{\textrm{\scriptsize{KS}}}_v(\textbf{r})\psi^{\textrm{\scriptsize{KS}}}_v(\textbf{r}')\\
&=&\sum_{\alpha\beta} \phi_\alpha(\textbf{r})P^{\{v\}\alpha\beta}\phi_\beta(\textbf{r}')
\end{eqnarray}
where $\{\phi_\alpha(\textbf{r})\}$ denotes a set of \emph{in situ} optimised\cite{onetep} localised atom-centered support functions referred to as non-orthogonal generalised Wannier functions (NGWFS)\cite{psinc} and $\textbf{P}^{\{v\}}$ is the single particle density matrix in the representation of those NGWFs. Linear scaling of computational cost with system size is then obtained by exploiting the fact that the ground-state density matrix decays exponentially for any system with a band gap, causing $\textbf{P}^{\{v\}}$ to be sparse for sufficiently large system size\cite{exponential1,exponential2}.

The \emph{in situ} optimisation of the support functions $\{\phi_\alpha\}$ means that only a minimal number of functions is required to accurately span the occupied subspace, but the unoccupied subspace is generally badly represented\cite{onetep_cond}. This issue is overcome by optimising a second set of NGWFs $\{\chi_\alpha\}$ for a low energy subset of the unoccupied subspace that is represented by the effective density matrix $\textbf{P}^{\{c\}}$\cite{onetep_cond}. It has been demonstrated\cite{tddft_onetep} that the compact sets of support functions $\{\phi_\alpha\}$ and $\{\chi_\beta\}$ provide a very good representation for low energy excited states as calculated in the TDA. Defining an effective response density $\rho^{\{1\}}(\textbf{r})$ associated with a TDA eigenvector $\textbf{X}$ such that 
\begin{eqnarray} \nonumber
\rho^{\{1\}}(\textbf{r})&=&\sum_{c,v}\psi^{\textrm{\scriptsize{KS}}}_c(\textbf{r}) \textrm{X}_{cv} \psi^{\textrm{\scriptsize{KS}}}_v(\textbf{r})\\
&=&\sum_{\alpha\beta}\chi_\alpha(\textbf{r})P^{\{1\}\alpha\beta}\phi_\beta(\textbf{r})
\end{eqnarray}
it becomes clear that the effective response density matrix $\textbf{P}^{\{1\}}$ is the representation of $\textbf{X}$ in mixed unoccupied-occupied NGWF space\cite{tddft_onetep}. The matrix-vector product $\textbf{f}=\textbf{A}\textbf{X}$ can then be directly constructed in NGWF space as 
\begin{eqnarray} \nonumber
\textbf{f}_{\textrm{\scriptsize{TDA}}}^{\chi\phi}&=&\textbf{P}^{\{c\}}\textbf{H}^\chi\textbf{P}^{\{1\}}-\textbf{P}^{\{1\}}\textbf{H}^\phi \textbf{P}^{\{v\}} \\
&&+\textbf{P}^{\{c\}}\left(\textbf{V}^{\{1\}\chi\phi}_{\textrm{\scriptsize{SCF}}}\left[\textbf{P}^{\{1\}}\right]\right)\textbf{P}^{\{v\}}.
\label{eqn:operator_TDA}
\end{eqnarray}
Here, $\textbf{H}^\chi$ and $\textbf{H}^\phi$ denote the ground state Kohn-Sham Hamiltonian in the $\{\chi_\alpha\}$ and $\{\phi_\beta\}$ representation respectively. $\textbf{V}^{\{1\}\chi\phi}_{\textrm{\scriptsize{SCF}}}\left[\textbf{P}^{\{1\}}\right]$ is the self-consistent field response of the system due to a perturbation $\rho^{\{1\}}(\textbf{r})$ in the ground state density\cite{tddft_onetep} and is the result of $\textbf{X}$ acting on the coupling matrix $\textbf{K}$ in mixed unoccupied-occupied NGWF space. Note that $\textbf{f}_{\textrm{\scriptsize{TDA}}}^{\chi\phi}$ represents a contravariant tensor quantity and has to be multiplied by $\textbf{S}^\chi$ and $\textbf{S}^{\phi}$ from the left and right respectively to obtain a covariant quantity (see \cite{tddft_onetep} for further details). The lowest excitation energy of the system can then be written as
\begin{equation}
\omega^{\textrm{\scriptsize{TDA}}}_\textrm{\scriptsize{min}}=\min_{\textbf{P}^{\{1\}}}\Omega_{\textrm{\scriptsize{TDA}}}\left[\textbf{P}^{\{1\}} \right]=\frac{\textrm{Tr}\left[\textbf{P}^{\{1\}\dagger}\textbf{S}^\chi\textbf{f}^{\chi\phi}\textbf{S}^\phi\right]}{\textrm{Tr}\left[\textbf{P}^{\{1\}\dagger}\textbf{S}^\chi\textbf{P}^{\{1\}}\textbf{S}^\phi\right]}
\label{eqn:functional_TDA}
\end{equation}
where $\textbf{S}^\chi$ and $\textbf{S}^\phi$ denote the overlap matrices of the $\{\chi_\alpha\}$ and $\{\phi_\beta\}$ NGWF representation respectively. Higher excited states can be obtained from the same variational principle by enforcing an orthogonality constraint between all excited states. If all involved density matrices $\textbf{P}^{\{1\}}$, $\textbf{P}^{\{v\}}$ and $\textbf{P}^{\{c\}}$ can be treated as sparse for sufficiently large system size\footnote{See the discussion in \cite{tddft_onetep} regarding the question when a truncation of $\textbf{P}^{\{1\}}$ is strictly justified}, evaluating Eqs.~\ref{eqn:operator_TDA} and \ref{eqn:functional_TDA} scales as $\mathcal{O}(N)$ with system size and $\omega^{\textrm{\scriptsize{TDA}}}_\textrm{\scriptsize{min}}$ can be computed in linear-scaling effort using standard iterative approaches. 

The above formulation yields accurate excitation energies if $\textbf{X}$ is well-represented by the low energy subset of unoccupied states for which $\textbf{P}^{\{c\}}$ is optimised. However, in many scenarios it is desirable to include higher energy unoccupied states in an approximate manner to achieve convergence\cite{tddft_onetep,jian_hao}. One straightforward way of doing so is to introduce the joint unoccupied-occupied representation $\{\varphi_\alpha\}=\{\phi_\beta\}\oplus\{\chi_\gamma\}$ and to redefine $\textbf{P}^{\{c\}}$ via a projector onto the entire unoccupied subspace representable by $\{\varphi_\alpha\}$\cite{jian_hao}:
\begin{equation}
\textbf{P}^{\{c\}}=\left(\textbf{S}^\varphi\right)^{-1}-\left(\textbf{S}^\varphi\right)^{-1}\textbf{S}^{\varphi\phi}\textbf{P}^{\{v\}}\textbf{S}^{\phi\varphi}\left(\textbf{S}^\varphi\right)^{-1}.
\label{eqn:proj}
\end{equation}
Here, the elements of $\textbf{S}^{\varphi\phi}$ are given by ${S}^{\varphi\phi}_{\alpha\beta}=\langle\varphi_\alpha|\phi_\beta \rangle$. While using the joint representation for the unoccupied subspace instead of $\{\chi_\alpha\}$ roughly doubles the computational cost compared to using $\{\chi_\alpha\}$, it yields consistently good TDDFT excitation energies\cite{jian_hao,thesis_tim} and will be used throughout in Sec.~\ref{sec:results}. However, for the purpose of clarity in outlining the linear-scaling TDDFT algorithm in Sec.~\ref{sec:full_tddft}, we shall use $\{\chi_\alpha\}$ to label the unoccupied space, noting that, if desired, the representation can be replaced by $\{\varphi_\beta\}$ and the projector of Eq.~\ref{eqn:proj}. 

\section{Full TDDFT in $\mathtt{ONETEP}$}
\label{sec:full_tddft}
In this section, we will outline a conjugate gradient algorithm to compute the lowest $N_\omega$ excited states of the full TDDFT equation. While other iterative eigensolvers like the Lanczos and Davidson algorithms and multishift methods have been applied to this problem, both in the framework of standard cubic scaling\cite{turbo,giustino} and $\mathcal{O}(N)$\cite{tretiak_tddft} approaches, the conjugate gradient method is chosen here for both its good performance in the TDA\cite{tddft_onetep} and its low memory requirements suitable for large scale applications. A special focus will be put on an effective preconditioning scheme, as well as the linear-scaling properties of the algorithm. As in the previous section, the discussion is limited to semi-local exchange-correlation functionals only. 

\subsection{The Tsiper functional in mixed $\{\chi_\alpha\}$-$\{\phi_\beta\}$ NGWF space}
\label{sec:operator}
The key to obtaining a linear-scaling implementation of the full TDDFT equation is to rewrite the Tsiper functional in the same compact, localised NGWF representation that has been used to rewrite the Tamm-Dancoff functional in Eq.~\ref{eqn:functional_TDA}. For this purpose, we define $\underline{\textbf{f}}_{\textrm{\scriptsize{Tsip}}}$ as
\begin{equation}
\underline{\textbf{f}}_{\textrm{\scriptsize{Tsip}}}=\begin{pmatrix} \textbf{f}_{\textrm{\scriptsize{p}}} \\ \textbf{f}_{\textrm{\scriptsize{q}}} \end{pmatrix}=\begin{pmatrix} \textbf{A}-\textbf{B} & \textbf{0} \\ \textbf{0} & \textbf{A}+\textbf{B} \end{pmatrix} \begin{pmatrix}\textbf{p} \\ \textbf{q} \end{pmatrix}=\begin{pmatrix} \textbf{A}\textbf{p}-\textbf{B}\textbf{p} \\ \textbf{A}\textbf{q}+\textbf{B}\textbf{q} \end{pmatrix}
\end{equation}
Following analogous steps to the derivation of the linear-scaling solution to the TDA eigenvalue equation\cite{tddft_onetep}, we define the effective response density matrices $\textbf{P}^{\{p\}}$ and $\textbf{P}^{\{q\}}$ and response densities $\rho^{\{p\}}(\textbf{r})$ and $\rho^{\{q\}}(\textbf{r})$ such that
\begin{eqnarray} \nonumber
\rho^{\{p\}}(\textbf{r})&=&\sum_{c,v}\psi^{\textrm{\scriptsize{KS}}}_c(\textbf{r}) \textrm{p}_{cv} \psi^{\textrm{\scriptsize{KS}}}_v(\textbf{r}) \\
&=&\sum_{\alpha\beta}\chi_\alpha(\textbf{r})P^{\{p\}\alpha\beta}\phi_\beta(\textbf{r})
\end{eqnarray}
with $\rho^{\{q\}}(\textbf{r})$ following an analogous definition. Thus $\textbf{P}^{\{p\}}$ and $\textbf{P}^{\{q\}}$ are the matrices $\textbf{p}$ and $\textbf{q}$ in $\{\chi_\alpha\}$-$\{\phi_\beta\}$ NGWF representation. Just like $\textbf{P}^{\{1\}}$\cite{tddft_onetep}, $\textbf{P}^{\{p\}}$ and $\textbf{P}^{\{q\}}$ have to follow an effective invariance constraint in order to be valid response density matrices, which originates from the orthogonality between the unoccupied and occupied Kohn-Sham spaces in which $\textbf{p}$ and $\textbf{q}$ are represented. For $\textbf{P}^{\{p\}}$, the invariance constraint can be written as  
\begin{equation}
\textbf{P}^{\{p\}'}=\textbf{P}^{\{c\}}\textbf{S}^\chi\textbf{P}^{\{p\}}\textbf{S}^\phi\textbf{P}^{\{v\}}=\textbf{P}^{\{p\}},
\label{eqn:invariance}
\end{equation}
with an identical statement for $\textbf{P}^{\{q\}}$.

From Sec.~\ref{sec:TDA}, the action of $\textbf{A}$ acting on some vector $\textbf{X}$ written in $\{\chi_\alpha\}$-$\{\phi_\beta\}$ NGWF space is already known. Using Eq.~\ref{eqn:operator_TDA}, it is straightforward to rewrite $\underline{\textbf{f}}_{\textrm{\scriptsize{Tsip}}}$ in $\{\chi_\alpha\}$ and $\{\phi_\beta\}$ representation such that 
\begin{eqnarray} \nonumber
\underline{\textbf{f}}_{\textrm{\scriptsize{Tsip}}}^{\chi\phi}&=&\begin{pmatrix} \textbf{f}^{\chi\phi}_{\{p\}}\\ \textbf{f}^{\chi\phi}_{\{q\}} \end{pmatrix}=\begin{pmatrix} \textbf{P}^{\{c\}}\textbf{H}^\chi\textbf{P}^{\{p\}}-\textbf{P}^{\{p\}}\textbf{H}^\phi \textbf{P}^{\{v\}} \\ \textbf{P}^{\{c\}}\textbf{H}^\chi\textbf{P}^{\{q\}}-\textbf{P}^{\{q\}}\textbf{H}^\phi \textbf{P}^{\{v\}}\end{pmatrix} \\ &+&\begin{pmatrix} \textbf{0} \\ 2\textbf{P}^{\{c\}}\left(\textbf{V}^{\{1\}\chi\phi}_{\textrm{\scriptsize{SCF}}}\left[\textbf{P}^{\{q\}}\right]\right)\textbf{P}^{\{v\}}\end{pmatrix}.
\label{eqn:operator_RPA}
\end{eqnarray}
The advantage of the reformulation of the Tsiper functional in terms of $\textbf{p}$ and $\textbf{q}$ now becomes apparent. In order to evaluate Eq.~\ref{eqn:operator_RPA} for any semi-local exchange-correlation functional, it is sufficient to evaluate $\textbf{V}^{\{1\}\chi\phi}_{\textrm{\scriptsize{SCF}}}$ only once. Since calculating $\textbf{V}^{\{1\}\chi\phi}_{\textrm{\scriptsize{SCF}}}$ is generally the most expensive part of applying the TDDFT operator, it follows that computing Eq.~\ref{eqn:operator_RPA} is not significantly more expensive than evaluating Eq.~\ref{eqn:operator_TDA}, suggesting that a full solution of to the TDDFT equations can be of similar computational complexity to the solution to the TDA\cite{scuseria}. 

Using Eq.~\ref{eqn:operator_RPA}, the lowest excitation energy of the system as specified by the Tsiper functional (\ref{eqn:tsiper_func}) can then be rewritten in $\{\chi_\alpha\}$ and $\{\phi_\beta\}$ representation
\begin{eqnarray} \nonumber
\omega_\textrm{\scriptsize{min}}&=& \min_{\left\{ \textbf{P}^{\{p\}},  \textbf{P}^{\{q\}} \right\}} \left\{  \frac{\textrm{Tr}\left[\textbf{P}^{\{p\}\dagger}\textbf{S}^\chi\textbf{f}^{\chi\phi}_{\{p\}} \textbf{S}^\phi \right]}{2\left|\textrm{Tr}\left[\textbf{P}^{\{p\}\dagger}\textbf{S}^\chi \textbf{P}^{\{q\}}\textbf{S}^\phi \right]\right|} \right. \\ &&+ \left.  \frac{\textrm{Tr}\left[\textbf{P}^{\{q\}\dagger}\textbf{S}^\chi\textbf{f}^{\chi\phi}_{\{q\}} \textbf{S}^\phi \right]}{2\left|\textrm{Tr}\left[\textbf{P}^{\{p\}\dagger}\textbf{S}^\chi \textbf{P}^{\{q\}}\textbf{S}^\phi \right]\right|} \right\}
\label{eqn:functional_RPA}
\end{eqnarray}
where the minimisation is carried out under the normalisation constraint
\begin{equation}
\textrm{Tr}\left[\textbf{P}^{\{p\}\dagger}\textbf{S}^\chi \textbf{P}^{\{q\}}\textbf{S}^\phi \right]=1.
\end{equation}
Specifying the normalisation constraint allows us to drop the absolute value from the denominator of the Tsiper functional when computing the gradient of Eq.~\ref{eqn:functional_RPA} with respect to changes in $\begin{pmatrix} \textbf{P}^{\{p\}} & \textbf{P}^{\{q\}} \end{pmatrix}$. Using Eq.~\ref{eqn:operator_RPA} the contravariant gradient of the Tsiper functional can be, in close analogy to the gradient in the TDA (\ref{eqn:grad_TDA}), written as
\begin{eqnarray} \nonumber
\underline{\textbf{g}}^{\chi\phi}_{\textrm{\scriptsize{Tsip}}}&=&\begin{pmatrix} \textbf{g}^{\chi\phi}_{\{p\}}\\ \textbf{g}^{\chi\phi}_{\{q\}} \end{pmatrix}=\begin{pmatrix}\textbf{f}^{\chi\phi}_{\{p\}}\\\textbf{f}^{\chi\phi}_{\{q\}} \end{pmatrix}-\left\{\textrm{Tr}\left[\textbf{P}^{\{p\}\dagger}\textbf{S}^\chi\textbf{f}^{\chi\phi}_{\{p\}} \textbf{S}^\phi \right]\right. \\ &&+\left.\textrm{Tr}\left[\textbf{P}^{\{q\}\dagger}\textbf{S}^\chi\textbf{f}^{\chi\phi}_{\{q\}} \textbf{S}^\phi \right] \right\} \begin{pmatrix}\textbf{P}^{\{q\}} \\ \textbf{P}^{\{p\}} \end{pmatrix}.
\label{eqn:tsip_grad}
\end{eqnarray}
This result exploits the fact that $\textbf{P}^{\{p\}}$ and $\textbf{P}^{\{q\}}$ follow the normalisation constraint. The above gradient can be used as a steepest-descent search direction in a conjugate gradient algorithm\cite{supplementary}. Note however, that since the TDDFT operator in the TDA can be computed in linear-scaling effort\cite{tddft_onetep}, the evaluation of both $\underline{\textbf{f}}_{\textrm{\scriptsize{Tsip}}}^{\chi\phi}$ and $\underline{\textbf{g}}^{\chi\phi}_{\textrm{\scriptsize{Tsip}}}$ also scales fully linearly with system size, as long as all involved density matrices $\textbf{P}^{\{c\}}$, $\textbf{P}^{\{v\}}$, $\textbf{P}^{\{p\}}$ and $\textbf{P}^{\{q\}}$ can be treated as sparse for sufficiently large system sizes. Furthermore, it is worth pointing out that for any matrix pair $\textbf{P}^{\{p\}}$ and $\textbf{P}^{\{q\}}$ obeying the invariance constraint of Eq.~\ref{eqn:invariance}, the gradient $\underline{\textbf{g}}^{\chi\phi}_{\textrm{\scriptsize{Tsip}}}$ follows the invariance constraint by construction. This condition is vital as it means that any pair of trial response density matrices updated with a search direction derived from $\underline{\textbf{g}}^{\chi\phi}_{\textrm{\scriptsize{Tsip}}}$ will also obey the appropriate invariance constraint by construction. 

It should be noted, that once a truncation of the density matrices $\textbf{P}^{\{q\}}$ and $\textbf{P}^{\{q\}}$ is introduced, the invariance constraint of Eqn. \ref{eqn:invariance} can only hold approximately. While it is possible to iteratively apply the projection at the end of each conjugate gradient step until some measure of the violation of the invariance constraint is kept below a certain threshold\cite{tddft_onetep, thesis_tim}, in practice this can lead to convergence problems as it destroys the variational nature of the algorithm presented here. This problem is overcome by introducing a set of auxiliary density matrices $\textbf{L}^{\{q\}}$ and $\textbf{L}^{\{q\}}$\cite{thesis_tim} in the spirit of ground state linear-scaling approaches\cite{LNV}. The auxiliary matrices can be arbitrarily truncated and are used to define the real density matrices $\textbf{P}^{\{q\}}$ and $\textbf{P}^{\{q\}}$ used in the algorithm via
\begin{equation}
\textbf{P}^{\{p\}}=\textbf{P}^{\{c\}}\textbf{S}^\chi\textbf{L}^{\{p\}}\textbf{S}^\phi\textbf{P}^{\{v\}}
\label{eqn:auxiliary}
\end{equation}
at every step of the calculation. While this scheme comes at a computational cost as $\textbf{P}^{\{p\}}$ is less sparse than $\textbf{L}^{\{p\}}$, it guarantees that every point in the algorithm, $\textbf{P}^{\{q\}}$ and $\textbf{P}^{\{q\}}$ fulfil their respective invariance constraints to the degree that $\textbf{P}^{\{v\}}$ and $\textbf{P}^{\{c\}}$ fulfil their respective idempotency constraint\cite{thesis_tim}. Since linear-scaling DFT calculations employ a number of techniques that ensure that for sensible truncation schemes, $\textbf{P}^{\{v\}}$ and $\textbf{P}^{\{c\}}$ retain near-idempotency\cite{LNV}, the above scheme yields a robust convergence of TDDFT calculations when truncations are applied to the respective response density matrices.

\subsection{Preconditioning}
\label{sec:precond}
The TDDFT eigenvalue problem generally has a large condition number associated with it, causing iterative eigensolvers to show a relatively slow convergence. This is most easily appreciated by considering that the elements of the coupling matrix $\textbf{K}$ are generally small compared to the diagonal elements of Kohn-Sham eigenvalue differences and the condition number of the full TDDFT matrix is reasonably well approximated by the condition number of $\begin{pmatrix} \textbf{D} & \textbf{0} \\ \textbf{0} &\textbf{D} \end{pmatrix}$, where the elements of the block matrix $\textbf{D}$ are given by
\begin{equation}
D_{cv,c'v'}=\delta_{cc'}\delta_{vv'}(\epsilon^{\textrm{\scriptsize{KS}}}_{c'}-\epsilon^{\textrm{\scriptsize{KS}}}_{v'}) .
\end{equation}
Clearly, $\textbf{D}$ has a condition number that is much larger than 1, resulting in relatively slow convergence of iterative eigensolvers. 

For these reasons, it has long been appreciated that $\begin{pmatrix} \textbf{D} & \textbf{0} \\ \textbf{0} &\textbf{D} \end{pmatrix}$ should form an efficient preconditioner for the full TDDFT eigenvalue problem. However, applying the preconditioner requires the computation of $\textbf{D}^{-1}$, which can only easily be constructed in a Kohn-Sham eigenstate representation where $\textbf{D}$ is diagonal. In linear-scaling TDDFT, a diagonal representation of $\textbf{D}$ is not available and the explicit construction of $\textbf{D}^{-1}$ via matrix inversion is undesirable.  

In order to obtain a linear-scaling preconditioner, we consider the Tsiper functional of Eq.~\ref{eqn:tsiper_func} when preconditioned from the left:
\begin{eqnarray} \nonumber
 &&\Omega_{\textrm{\scriptsize{Tsip}}}(\textbf{p},\textbf{q})= \\ &&\frac{\begin{pmatrix}\textbf{p}^\dagger & \textbf{q}^\dagger \end{pmatrix}\begin{pmatrix}\textbf{D}^{-1} \left(\textbf{A}-\textbf{B}\right) & \textbf{0} \\ \textbf{0} &\textbf{D}^{-1} \left(\textbf{A}+\textbf{B} \right)\end{pmatrix} \begin{pmatrix}\textbf{p} \\ \textbf{q} \end{pmatrix}}{\left | \begin{pmatrix}\textbf{p}^\dagger & \textbf{q}^\dagger \end{pmatrix}  \begin{pmatrix} \textbf{0} & \textbf{D}^{-1} \\ \textbf{D}^{-1} & \textbf{0} \end{pmatrix} \begin{pmatrix}\textbf{p} \\ \textbf{q} \end{pmatrix} \right |}
\end{eqnarray}
While the action of $\textbf{D}^{-1}$ on a matrix cannot be straightforwardly constructed in $\{\chi_\alpha\}$-$\{\phi_\beta\}$ NGWF space, the action of $\textbf{D}$ is trivially known. Denoting $\underline{\textbf{G}}^{\chi\phi}_{\textrm{\scriptsize{Tsip}}}=\begin{pmatrix}\textbf{G}^{\chi\phi}_{\{p\}}\\\textbf{G}^{\chi\phi}_{\{q\}} \end{pmatrix}$ as the preconditioned version of the gradient $\underline{\textbf{g}}^{\chi\phi}_{\textrm{\scriptsize{Tsip}}}$ of Eq.~\ref{eqn:tsip_grad}, it can be seen that applying the preconditioner $\textbf{D}^{-1}$ to the Tsiper functional is equivalent to solving the linear system
\begin{equation}
\begin{pmatrix} \textbf{P}^{\{c\}}\textbf{H}^\chi\textbf{G}^{\chi\phi}_{\{p\}}-\textbf{G}^{\chi\phi}_{\{p\}}\textbf{H}^\phi \textbf{P}^{\{v\}}\\ \textbf{P}^{\{c\}}\textbf{H}^\chi\textbf{G}^{\chi\phi}_{\{q\}}-\textbf{G}^{\chi\phi}_{\{q\}}\textbf{H}^\phi \textbf{P}^{\{v\}}\end{pmatrix}=\underline{\textbf{g}}^{\chi\phi}_{\textrm{\scriptsize{Tsip}}}
\label{eqn:precond}
\end{equation}
for $\underline{\textbf{G}}^{\chi\phi}_{\textrm{\scriptsize{Tsip}}}$. This linear system can be solved iteratively to a chosen degree of numerical accuracy using a standard conjugate gradient algorithm (see algorithm 2 detailed in \cite{supplementary}). While applying the preconditioner in NGWF space is therefore not as straightforward as in Kohn-Sham space, it scales fully linearly with system size and only requires a number of comparatively inexpensive matrix-matrix multiplications. If the computational overhead from solving the linear system in every step of the conjugate gradient calculation is significantly less than the time saved by constructing $\underline{\textbf{g}}^{\chi\phi}_{\textrm{\scriptsize{Tsip}}}$ fewer times due to a faster convergence rate, the preconditioning proposed here becomes highly efficient.  This point will be addressed in more detail in Sec.~\ref{sec:azo}.

\subsection{Optimising multiple excited states}
\label{sec:multiple}
In most situations, we are not interested in converging only the lowest excited state of the full TDDFT equation, but rather the subspace spanning the lowest $N_\omega$ excitations $\{\omega_i\}$. Following the considerations in Sec. \ref{sec:tsiper} regarding the convergence of higher excited states, we introduce a set of $N_\omega$ TDDFT trial vectors $\{\underline{\textbf{P}}^{\textrm{\scriptsize{Tsip}}}_i;\,i=1,\cdots N_\omega\}$, where
\begin{equation}
\underline{\textbf{P}}^{\textrm{\scriptsize{Tsip}}}_i=\begin{pmatrix}\textbf{P}^{\{p\}}_i \\ \textbf{P}^{\{q\}}_i \end{pmatrix}.
\end{equation}
In order to span the appropriate subspace, $\{\underline{\textbf{P}}^{\textrm{\scriptsize{Tsip}}}_i\}$ is required to follow an orthonormality condition that is written as
\begin{equation}
\frac{1}{2}\left(\textrm{Tr}\left[\textbf{P}_i^{\{p\}\dagger}\textbf{S}^\chi \textbf{P}_j^{\{q\}}\textbf{S}^\phi \right]+\textrm{Tr}\left[\textbf{P}_j^{\{p\}\dagger}\textbf{S}^\chi \textbf{P}_i^{\{q\}}\textbf{S}^\phi \right]\right)=\delta_{ij} .
\label{eqn:ortho}
\end{equation}
In the algorithm presented here, in close analogy to the way the Tamm-Dancoff eigenvalue problem is treated\cite{tddft_onetep}, the orthonormality condition is enforced using a Gram-Schmidt procedure. However, some additional care has to be taken, since the quantity $\textrm{Tr}\left[\textbf{P}_i^{\{p\}\dagger}\textbf{S}^\chi \textbf{P}_i^{\{q\}}\textbf{S}^\phi \right]$ is not required to be positive-semidefinite. Following the convention established in Ref.~\cite{tretiak_tddft}, prior to orthogonalising the set $\{\underline{\textbf{P}}^{\textrm{\scriptsize{Tsip}}}_i\}$, $\textrm{Tr}\left[\textbf{P}_i^{\{p\}\dagger}\textbf{S}^\chi \textbf{P}_i^{\{q\}}\textbf{S}^\phi \right]$ is computed for all $i$ and, if found negative, the vector $\underline{\textbf{P}}^{\textrm{\scriptsize{Tsip}}}_i$ is transformed according to
\begin{equation}
\begin{pmatrix}\textbf{P}^{\{p\}}_i \\ \textbf{P}^{\{q\}}_i \end{pmatrix} \rightarrow \begin{pmatrix}\textbf{P}^{\{p\}}_i \\ -\textbf{P}^{\{q\}}_i \end{pmatrix}.
\end{equation}

Once an orthonormal set $\{\underline{\textbf{P}}^{\textrm{\scriptsize{Tsip}}}_i;\,i=1,\cdots N_\omega\}$ has been obtained, it can be used to construct $\left\{ (\underline{\textbf{f}}_{\textrm{\scriptsize{Tsip}}}^{\chi\phi})_i;\,i=1,\cdots N_\omega  \right\}$ from Eq.~\ref{eqn:operator_RPA}. It is then possible to write down an effective functional for the sum of the lowest $N_\omega$ excitations of the system, such that
\begin{equation}
\min_{\left\{ \underline{\textbf{P}}^{\textrm{\scriptsize{Tsip}}}_i \right\}} \Omega_{\textrm{\scriptsize{Tsip}}}^{N_\omega}\left(\left\{ \underline{\textbf{P}}^{\textrm{\scriptsize{Tsip}}}_i\right\} \right) = \sum_i^{N_\omega} \omega_i
\label{eqn:min_multiple}
\end{equation}
where
\begin{eqnarray} \nonumber
\Omega_{\textrm{\scriptsize{Tsip}}}^{N_\omega}\left(\left\{ \underline{\textbf{P}}^{\textrm{\scriptsize{Tsip}}}_i\right\} \right)&=& \sum_i^{N_\omega} \left\{ \frac{\textrm{Tr}\left[\textbf{P}_i^{\{p\}\dagger}\textbf{S}^\chi\left(\textbf{f}^{\chi\phi}_{\{p\}}\right)_i \textbf{S}^\phi \right]}{2\left|\textrm{Tr}\left[\textbf{P}_i^{\{p\}\dagger}\textbf{S}^\chi \textbf{P}_i^{\{q\}}\textbf{S}^\phi \right]\right|} \right. \\ 
&&+\left.\frac{\textrm{Tr}\left[\textbf{P}_i^{\{q\}\dagger}\textbf{S}^\chi\left(\textbf{f}^{\chi\phi}_{\{q\}}\right)_i \textbf{S}^\phi \right]}{2\left|\textrm{Tr}\left[\textbf{P}_i^{\{p\}\dagger}\textbf{S}^\chi \textbf{P}_i^{\{q\}}\textbf{S}^\phi \right]\right|} \right\}.
\label{eqn:Tsiper_multiple}
\end{eqnarray}
and the minimisation in Eqn. \ref{eqn:min_multiple} is to be carried out under the effective orthonormality constraint placed on $\left\{ \underline{\textbf{P}}^{\textrm{\scriptsize{Tsip}}}_i \right\}$.

Differentiating the above expression with respect to $\underline{\textbf{P}}^{\textrm{\scriptsize{Tsip}}}_i$, it is possible to construct a contravariant gradient $\left(\underline{\textbf{g}}^{\chi\phi}_{\textrm{\scriptsize{Tsip}}}\right)_i$ that is orthogonal to all current subspace vectors $\{\underline{\textbf{P}}^{\textrm{\scriptsize{Tsip}}}_i\}$. The gradient can be written as
\begin{eqnarray} \nonumber
\left(\underline{\textbf{g}}^{\chi\phi}_{\textrm{\scriptsize{Tsip}}}\right)_i&=&\left(\underline{\textbf{f}}_{\textrm{\scriptsize{Tsip}}}^{\chi\phi}\right)_i-\sum_j^{N_\omega} \left\{ \frac{\textrm{Tr}\left[\textbf{P}_j^{\{p\}\dagger}\textbf{S}^\chi\left(\textbf{f}^{\chi\phi}_{\{p\}}\right)_i \textbf{S}^\phi \right]}{2\textrm{Tr}\left[\textbf{P}_j^{\{p\}\dagger}\textbf{S}^\chi \textbf{P}_j^{\{q\}}\textbf{S}^\phi\right]}\right. \\ &&+\left. \frac{\textrm{Tr}\left[\textbf{P}_j^{\{q\}\dagger}\textbf{S}^\chi\left(\textbf{f}^{\chi\phi}_{\{q\}}\right)_i \textbf{S}^\phi \right]}{2\textrm{Tr}\left[\textbf{P}_j^{\{p\}\dagger}\textbf{S}^\chi \textbf{P}_j^{\{q\}}\textbf{S}^\phi\right]} \right\}\begin{pmatrix}\textbf{P}^{\{q\}}_j \\ \textbf{P}^{\{p\}}_j  \end{pmatrix}.
\end{eqnarray}
The fact that $\left(\underline{\textbf{g}}^{\chi\phi}_{\textrm{\scriptsize{Tsip}}}\right)_i$ is orthogonal to all current subspace vectors $\{\underline{\textbf{P}}^{\textrm{\scriptsize{Tsip}}}_i\}$ is crucial for the correct performance of the algorithm. It ensures that when using $\left(\underline{\textbf{g}}^{\chi\phi}_{\textrm{\scriptsize{Tsip}}}\right)_i$ as a steepest descent direction to update $\underline{\textbf{P}}^{\textrm{\scriptsize{Tsip}}}_i$ such that
\begin{equation}
\underline{\textbf{P}}^{\textrm{\scriptsize{Tsip}}}_i\rightarrow \underline{\textbf{P}}^{\textrm{\scriptsize{Tsip}}}_i+\lambda \left(\underline{\textbf{g}}^{\chi\phi}_{\textrm{\scriptsize{Tsip}}}\right)_i
\end{equation}
for some given line step $\lambda$, there is no violation of the orthonormality constraint of Eq.~\ref{eqn:ortho} to first order in $\lambda$. 

The above outline contains all of the basic ingredients to construct a preconditioned conjugate gradient algorithm capable of solving for the lowest $N_\omega$ TDDFT excitations of the system in linear-scaling effort. The exact algorithm used in the implementation discussed in this work is adapted from \cite{peter_cg} for the purpose of solving the full TDDFT eigenvalue problem (See algorithm 1 in \cite{supplementary}). Here we limit ourselves to the comment that while minimising the Tsiper functional of Eq.~\ref{eqn:Tsiper_multiple} yields a set of vectors spanning the same subspace as the TDDFT eigenvectors corresponding to the lowest $N_\omega$ eigenvalues, these eigenvalues can only be obtained through a subspace diagonalisation that scales as $\mathcal{O}(N_\omega^3)$. However, since $N_\omega$ is generally taken to be small, this single cubic scaling step is not considered to be a bottleneck in any practical calculation. The subspace diagonalisation can be carried out by diagonalising the $N_\omega\times N_\omega$ dimensional matrix $\textbf{Q}$ with matrix elements
\begin{equation}
Q_{ij}=\frac{\textrm{Tr}\left[\textbf{P}_j^{\{p\}\dagger}\textbf{S}^\chi\left(\textbf{f}^{\chi\phi}_{\{p\}}\right)_i \textbf{S}^\phi \right]+\textrm{Tr}\left[\textbf{P}_j^{\{q\}\dagger}\textbf{S}^\chi\left(\textbf{f}^{\chi\phi}_{\{q\}}\right)_i \textbf{S}^\phi \right] }{2}.
\end{equation}

While the subspace diagonalisation only has to be carried out at the end of the calculation, the Gram-Schmidt orthonormalisation has an $\mathcal{O}(N_\omega^2)$ scaling associated with it and thus, like the TDA implementation\cite{tddft_onetep}, the algorithm only shows a linear-scaling with system size for a constant number of excitations. However, we point out that for many systems, especially the large pigment-protein complexes mentioned as the main focus of this work, low energy absorption properties are dominated by a small number of excitations of interest that stays constant with system size and for this class of systems, the method presented here allows for truly linear-scaling calculations. 

\section{Results}
\label{sec:results}
In this section we will present some of the strengths of the algorithm developed in Sec.~\ref{sec:full_tddft} and specifically address the question of whether full TDDFT produces significantly more accurate results compared to the TDA for a certain class of systems. We will first focus on $\emph{trans}$-azobenzene, a molecule small enough to be easily treatable with standard cubic-scaling implementations of TDDFT. After comparing the linear-scaling TDDFT approach presented here to these benchmark calculations we will discuss the importance of preconditioning the full TDDFT eigenvalue equation in order to speed up convergence. Section~\ref{sec:bcl} then aims at reproducing experimental results of bacteriochlorophyll $a$ in an organic solvent, addressing again the question of whether full TDDFT provides a significant advantage over the TDA in this system, as well as the influence of an explicit treatment of solvent molecules on the spectrum. Finally, it is demonstrated that the algorithm is capable of obtaining low energy excitations in linear-scaling effort by computing the excitations in system sizes of up to $\approx$ 7000 atoms.  

All calculations performed in this section are done using the PBE functional\cite{pbe}. Norm-conserving pesudopotentials\cite{pseudopot}, as well as the projection operator and the joint NGWF set of Eq.~\ref{eqn:proj} are used throughout for the $\mathtt{ONETEP}$ TDDFT calculations. 

\subsection{\emph{Trans}-azobenzene}
\label{sec:azo}
As a first test system, we choose \emph{trans}-azobenzene (C$_{12}$H$_{10}$N$_2$) as its moderate size allows for detailed benchmark comparisons to conventional TDDFT implementations showing an $O(N^3)$ scaling. Furthermore, the optical spectrum of this system has already been studied to a high degree of accuracy using the $GW$ approximation and the Bethe-Salpeter equation (BSE), where it was found that the TDA generated considerable errors compared to a full solution to the Bethe-Salpeter equation\cite{azobenzene_benchmark}. While no straightforward comparison can be drawn between the $GW$+BSE and TDDFT, the similar structure of the equations leads us to expect that TDDFT in the TDA and full TDDFT will also yield significantly different results for this system, making it an ideal test case. 

\begin{table}
\begin{centering}
\begin{tabular}{|c|c|c|c|c|}
\hline
  & $\mathtt{ONETEP}$ TDA & NWChem TDA & $\mathtt{ONETEP}$ RPA & NWChem RPA \\ \hline
1 & 2.233 & 2.192& 2.184& 2.149\\
2 & 3.520(0.047)& 3.524(0.091) &3.516(0.286) & 3.518(0.186) \\
3 & 3.536(0.001)& 3.546(0.001) & 3.489(0.002) & 3.499(0.003)\\
4 & 3.720(1.094)& 3.681(1.025)&  3.408(0.499)& 3.379(0.751)\\
5 & 3.822& 3.866&3.821 &3.865 \\
6 & 3.875(0.003)& 3.923(0.001) & 3.874 & 3.922 \\
7 & 4.234(0.001)& 4.268(0.001)& 4.229(0.001) & 4.262(0.001)\\
8 & 4.315& 4.305 &4.241 & 4.230 \\ \hline
\end{tabular}
\end{centering}
\caption{Lowest eight excited states of azobenzene, as computed with $\mathtt{ONETEP}$ and NWChem. Energies are given in eV, oscillator strengths are shown in brackets. States without specified oscillator strengths are dark. Where necessary, the states have been reordered according to their character, such that the order is the same as for the $\mathtt{ONETEP}$ TDA results. For the NWChem calculations, an aug-cc-pVTZ Gaussian basis set is used. Here, TDA denotes the Tamm-Dancoff approximation, while RPA is used to denote a solution to the full TDDFT equation.}
\label{tab:azo}
\end{table}

First the ionic positions of \emph{trans}-azobenzene are optimised\cite{onetep_force} in $\mathtt{ONETEP}$, after which the lowest eight excitations are computed both in the TDA and in full TDDFT. The TDA results are obtained using the algorithm introduced in Ref.~\cite{tddft_onetep}, while the full TDDFT results are computed using the algorithm introduced in this work. Calculations are performed using a kinetic energy cutoff of 1000~eV and a box size of $56.69\times56.69\times56.69$~\AA$^3$. In order to avoid any interaction between periodic images, the TDDFT calculations are carried out in open boundary conditions\cite{open_bc}. A minimal set of one NGWF per H and four NGWFs for each C and N atom is used both for $\{\chi_\alpha\}$ and $\{\phi_\beta\}$. A localisation radius of 10~$a_0$ is applied to $\{\phi_\beta\}$ representing the occupied subspace, while the localisation is relaxed to 13~$a_0$ for $\{\chi_\alpha\}$ in order to better represent the more delocalised unoccupied states. In some molecules, low energy excitations can be found that are of very delocalised Rydberg-state character and provide a challenge for localised basis set representations. However, tests performed using TDA have shown that these states can be systematically converged by increasing localisation radius of $\{\chi_\alpha\}$, and converged values are generally found to be in good agreement with results obtained from real-space methods\cite{tddft_onetep}. For the purpose of this work, we have performed a number of convergence tests, increasing the localisation radius to 18~$a_0$ and found the lowest eight excitations of alizarin to be well converged at the localisation radius of 13~$a_0$.

The TDDFT results from $\mathtt{ONETEP}$ are compared to results obtained using NWChem\cite{nwchem}. The NWChem calculations are performed using the same ionic positions as in $\mathtt{ONETEP}$ and an aug-cc-pVTZ Gaussian basis set containing diffuse functions that are designed to yield a good description of weakly bound unoccupied states\cite{basis_set}. In order to make the all-electron calculations more comparable to the pseudopotential calculations in $\mathtt{ONETEP}$, the Kohn-Sham states corresponding to the core electrons are excluded from the occupied subspace when calculating the TDDFT excitation energies. 

The results for the lowest eight excitations for the TDA and full TDDFT as calculated using $\mathtt{ONETEP}$ in comparison to the NWChem results can be found in Table \ref{tab:azo}. As can be seen, there is a generally good agreement between the $\mathtt{ONETEP}$ and the NWChem results, both in terms of the excitation energy and the oscillator strength. The largest discrepancy is found in the the sixth excited state, with the $\mathtt{ONETEP}$ results being 48~meV lower in energy for both the TDA and full TDDFT. Most other results show a significantly smaller discrepancy. 

Generally, the full TDDFT results compare very similarly to the NWChem benchmark as the TDA results, showing that the algorithm indeed performs correctly. Remaining discrepancies are most likely due to the all-electron treatment in NWChem compared to the pseudopotential treatment in $\mathtt{ONETEP}$, as well as basis set differences (see Ref.~\cite{tddft_onetep} for a detailed comparison between NWChem and $\mathtt{ONETEP}$ regarding TDA results). It is thus evident that the minimal NGWF representation is capable of obtaining excitation energies of a comparable quality to those from considerably larger Gaussian basis set representations, highlighting the advantages of using an $\emph{in situ}$ optimised representation when performing TDDFT calculations.  

\begin{figure}
\centering
\resizebox{0.48\textwidth}{!}{\includegraphics{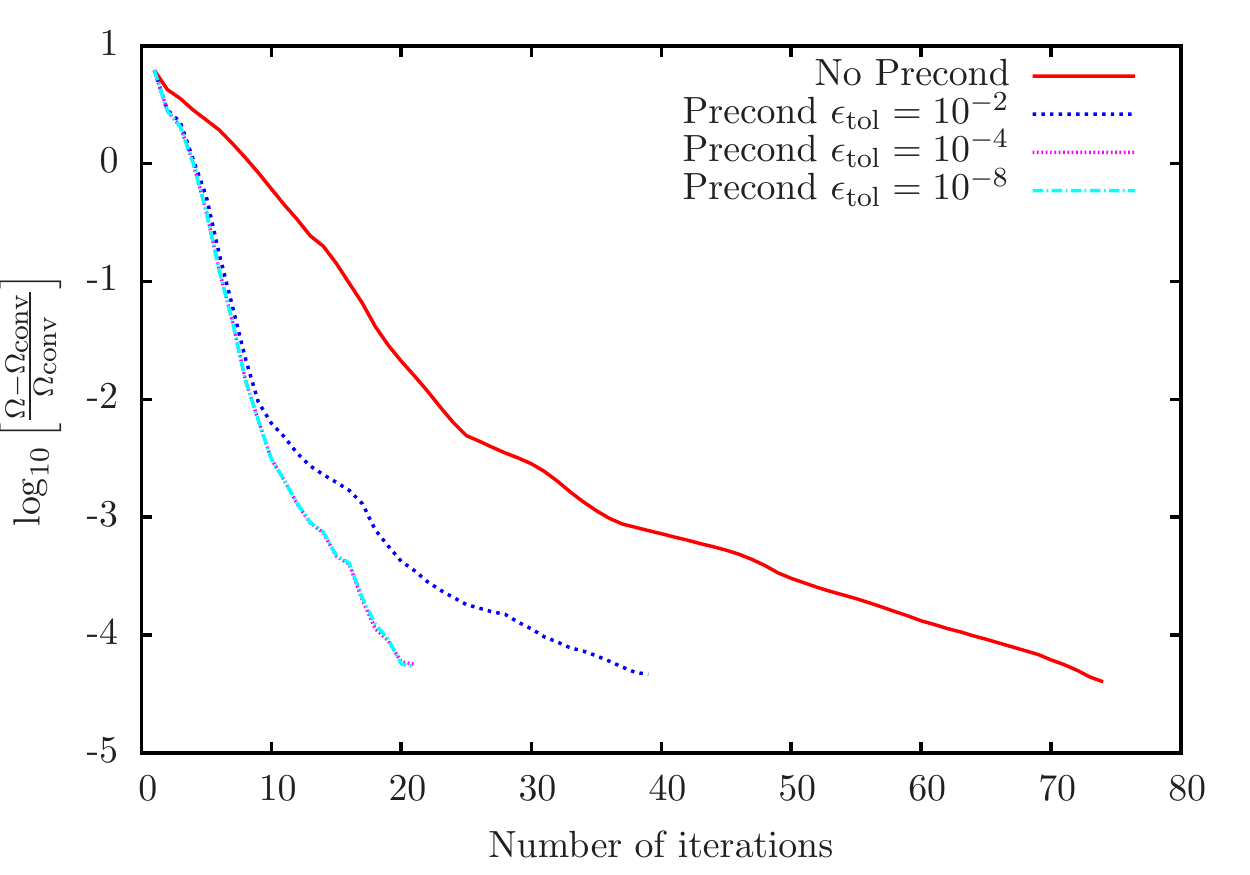}}
\caption{Convergence of the 2 lowest states of Azobenzene for different degrees of preconditioning applied. ``Precond $\epsilon_{\mathrm{tol}}$'' describes the tolerance to which the linear system (\ref{eqn:precond}) is solved in order to apply the preconditioning.}
\label{fig:precond}
\end{figure}

Regarding the impact of full TDDFT on the low energy excited states of \emph{trans}-azobenzene, it becomes apparent that when solving the full TDDFT equations, the dominant low energy excited state, which is mainly made up of a HOMO-1$\rightarrow$LUMO transition, decreases in energy by more than 0.31~eV as compared with the TDA results.
 The peak with the second highest oscillator strength, corresponding to a HOMO-2$\rightarrow$LUMO transition, stays almost constant in energy. In the full TDDFT, a significant part of spectral weight is shifted from the dominant peak to the second peak. This result is in good qualitative agreement with the $GW$+BSE results, where it was found that the TDA blue-shifts the dominant peak by 0.2~eV and overestimates its oscillator strength\cite{azobenzene_benchmark}. It is thus clear that solving the full TDDFT equation instead of the TDA can lead to significant changes in the computed optical spectrum, with bright peaks being shifted by tenths of electronvolts and a significant redistribution of spectral weight occouring. 

The small size of \emph{trans}-azobenzene makes it an ideal system verify the effectiveness of the preconditioning introduced in this work. For this purpose, the we compute the two lowest excited states of the system for a number of different convergence tolerances $\epsilon_{\mathrm{tol}}$ of the iterative preconditioner\cite{supplementary}, as well as the case when no preconditioning is applied to the conjugate gradient algorithm. The convergence rate of the two lowest excited states with respect to different levels of preconditioning applied can be found in Fig.~\ref{fig:precond}. Note that iteratively applying the preconditioner to a tolerance of $10^{-4}$ cuts the number of iterations needed to reach convergence by almost a factor of four compared to the case where no preconditioning is applied. This highlights the ill-conditioning of the TDDFT eigenvalue equation mentioned in section \ref{sec:precond} and shows that preconditioning is vital in achieving good convergence rates. However, note that even when the iterative preconditioning tolerance is set as high as $10^{-2}$, corresponding to only applying the preconditioner approximately at each iteration, the number of iterations needed to reach convergence is decreased by a factor of two. This finding is vital as solving the linear system of Eq.~\ref{eqn:precond} iteratively in each conjugate gradient step has a computational overhead associated with it, which can be minimised if the solution is only obtained approximately rather than to a high degree of numerical accuracy. It should be pointed out however, that the computational overhead with associated with the preconditioning is negligible for the system at hand, making up 0.17\% and 0.53\% of the total calculation time for $\epsilon_{\mathrm{tol}}=10^{-2}$ and  $\epsilon_{\mathrm{tol}}=10^{-4}$, which can again be attributed to the compact size of the NGWF representation. Furthermore, it is found that the total calculation time is reduced by a factor of 2.86 when comparing the unconditioned system to a preconditioned system with $\epsilon_{\mathrm{tol}}=10^{-4}$, thus showing that the preconditioner introduced here indeed leads to significant reductions in computational effort. 

We note that the convergence of the preconditioned system with the high convergence tolerance of $\epsilon_\textrm{\scriptsize{tol}}=10^{-2}$ closely follows the fast convergence of the tight tolerance results of $\epsilon_\textrm{\scriptsize{tol}}=10^{-8}$ for the first ten iterations. This suggests that the convergence tolerance of the preconditioner can be chosen adaptively. Starting off with a high tolerance for the first few iterations and tightening it closer to convergence has the potential to provide the ideal balance between reducing the computational overhead of the preconditioning and increasing the convergence rate. In conclusion it is demonstrated that the preconditioned conjugate gradient algorithm introduced in this work yields a very good agreement with existing TDDFT implementations and shows excellent convergence rates. 

\subsection{Bacteriochlorophyll}
\label{sec:bcl}
We now shift the focus to bacteriochlorophyll $a$ (MgN$_4$O$_6$C$_{55}$H$_{74}$), a chromophore that is of great interest in computational biology due to its role in light-harvesting complexes\cite{shim_fmo,neugebauer_fmo,danny_fmo}. Bacteriochlorophyll (BChl) is a medium-sized system still within the range of conventional cubic scaling TDDFT implementations. Therefore, the focus in this section is not to demonstrate the linear-scaling capabilities of the algorithm developed in this work but rather to address the question of whether full TDDFT yields a better description of the low energy absorption spectrum than the TDA when compared to experimental results. 

\begin{figure}
\centering
\resizebox{0.22\textwidth}{!}{\includegraphics[natwidth=5in,natheight=3.5in]{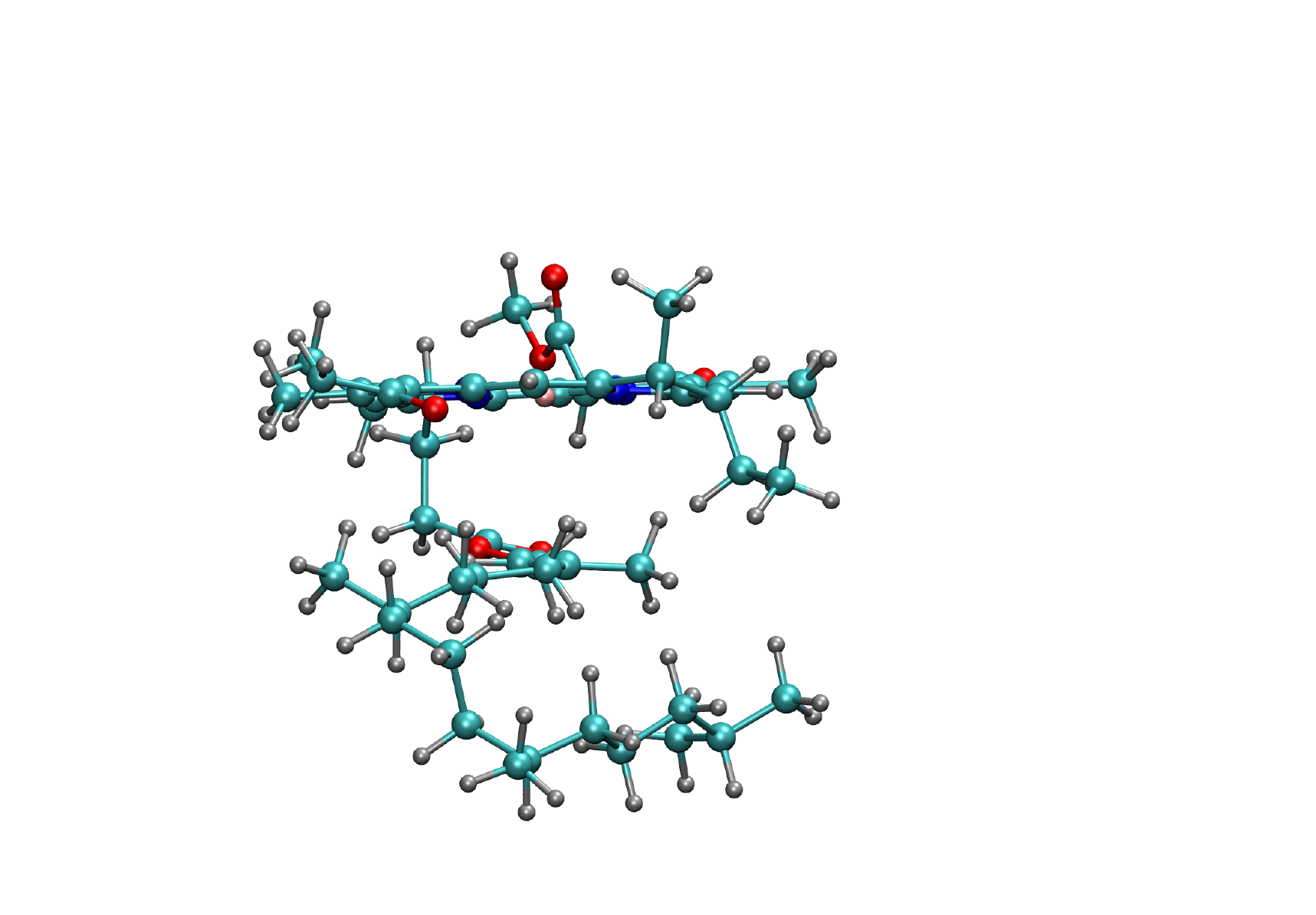}}
\resizebox{0.22\textwidth}{!}{\includegraphics[natwidth=5in,natheight=3.5in]{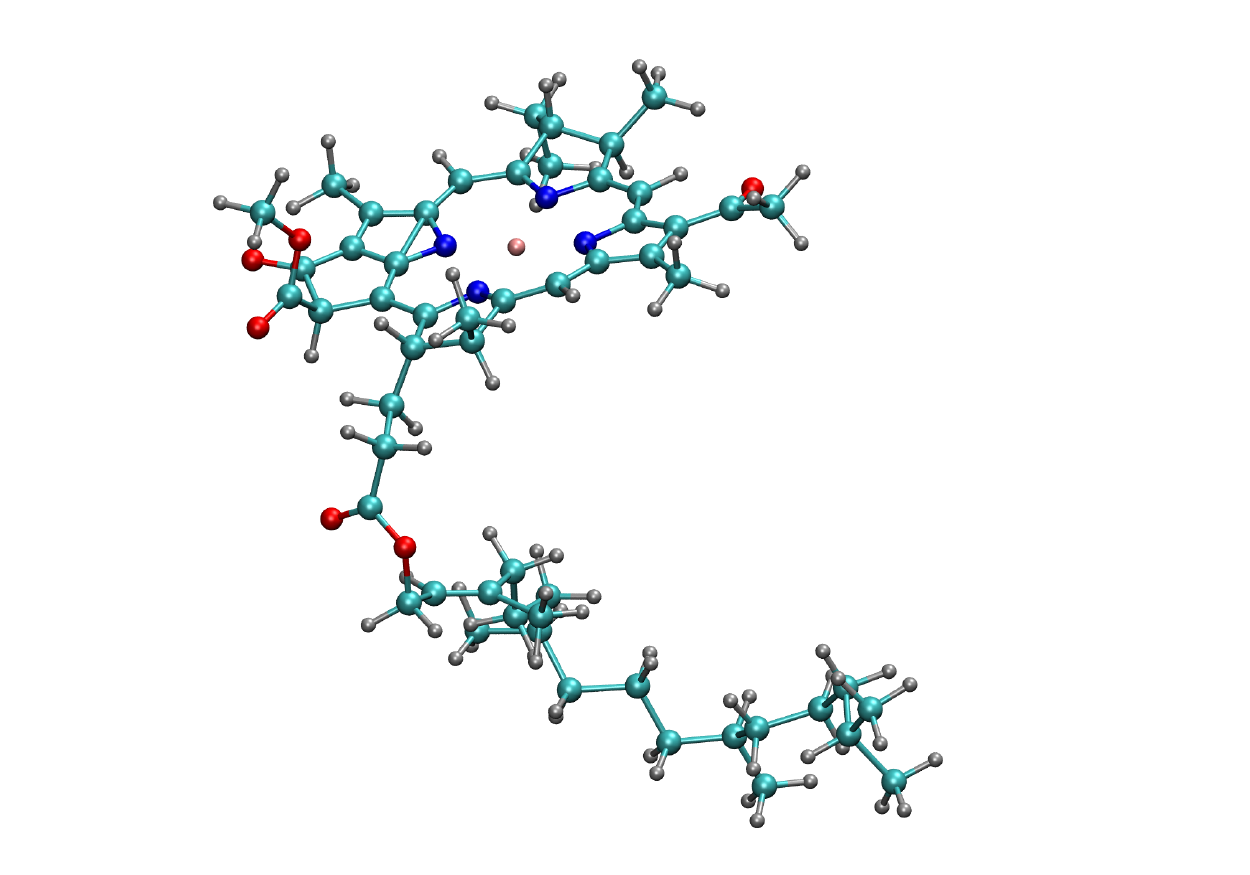}}
\caption{Structure of bacteriochlorophyll $a$, as obtained from a geometry optimisation in vacuum (left) and a single snapshot from an MD simulation in toluene (right). This figure was created using VMD\cite{vmd}.}
\label{fig:toluene_struct}
\end{figure}

All TDDFT calculations performed in this section are carried out using a minimal set of NGWFs for all atomic species involved and a localisation radius of 10~$a_0$ and 13~$a_0$ for the NGWFs representing the occupied and unoccupied space respectively. A kinetic energy cutoff of 1020~eV is used in all calculations. 

We aim to compare to the experimental results for the low energy absorption spectrum of BChl $a$ in a toluene organic solvent\cite{bchl_original_experiment,bchl_experiment}. As a first step, we take the atomic positions of BChl $a$ site~1 in the Fenna-Matthews-Olson complex (see Ref.~\cite{danny_fmo} for an explanation of how the input structure was obtained from X-ray diffraction results), optimise the atomic positions in vacuum and then calculate the TDA and full TDDFT spectra of the system within an implicit solvent model\cite{implicit_solvent}, where the static dielectric function $\epsilon$ is chosen to be 2.38 in order to match the dielectric function of toluene at room temperature. The final structure of BChl $a$ in vacuum can be found in Fig.~\ref{fig:toluene_struct}. As can be seen, the porphine ring is entirely flat in this configuration, while the alkane tail folds underneath the ring structure.  

\begin{figure}
\centering
\subfloat[Implicit solvent, vacuum structure\label{fig:bchl_implicit}]{\resizebox{0.48\textwidth}{!}{\includegraphics{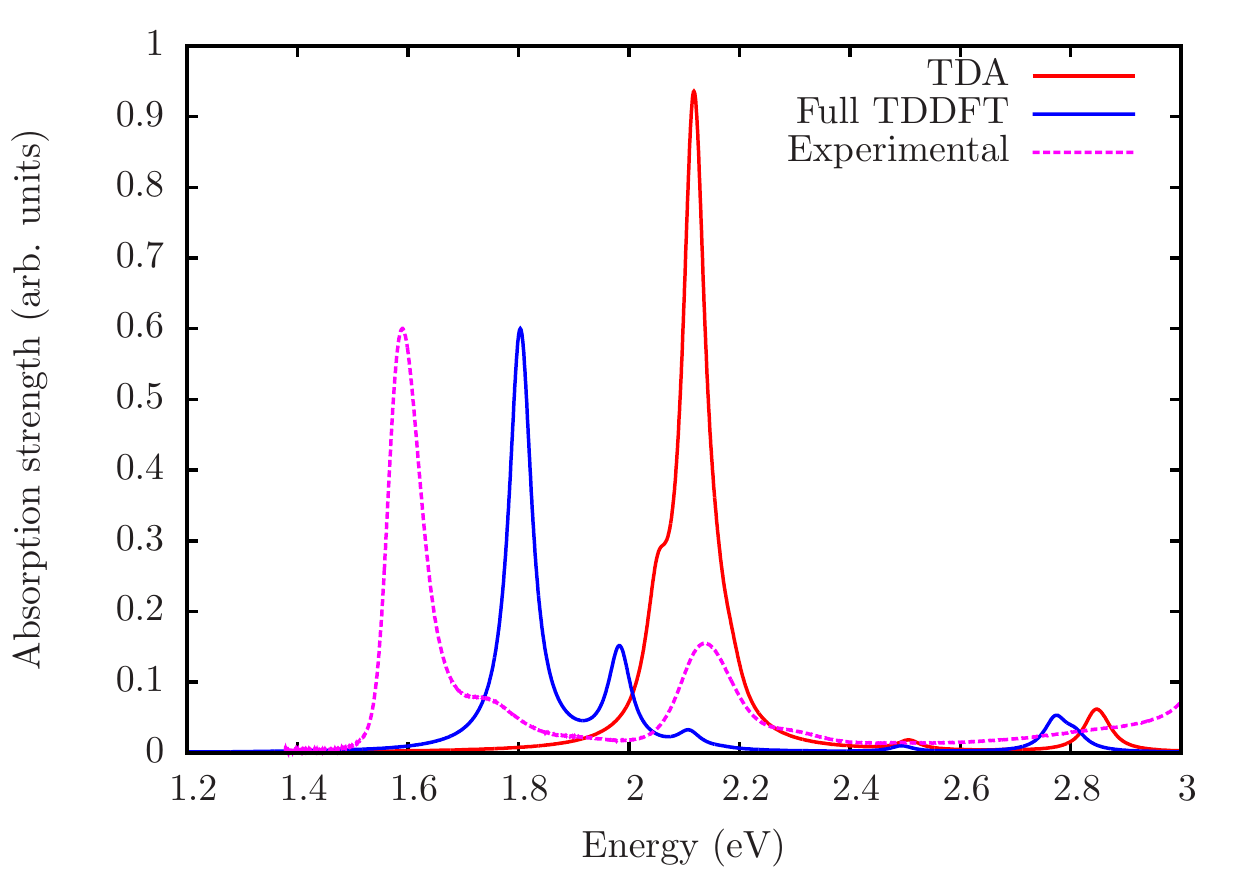}}}
\\
\subfloat[Explicit solvent, averaged snapshots \label{fig:bchl_explicit}]{\resizebox{0.48\textwidth}{!}{\includegraphics{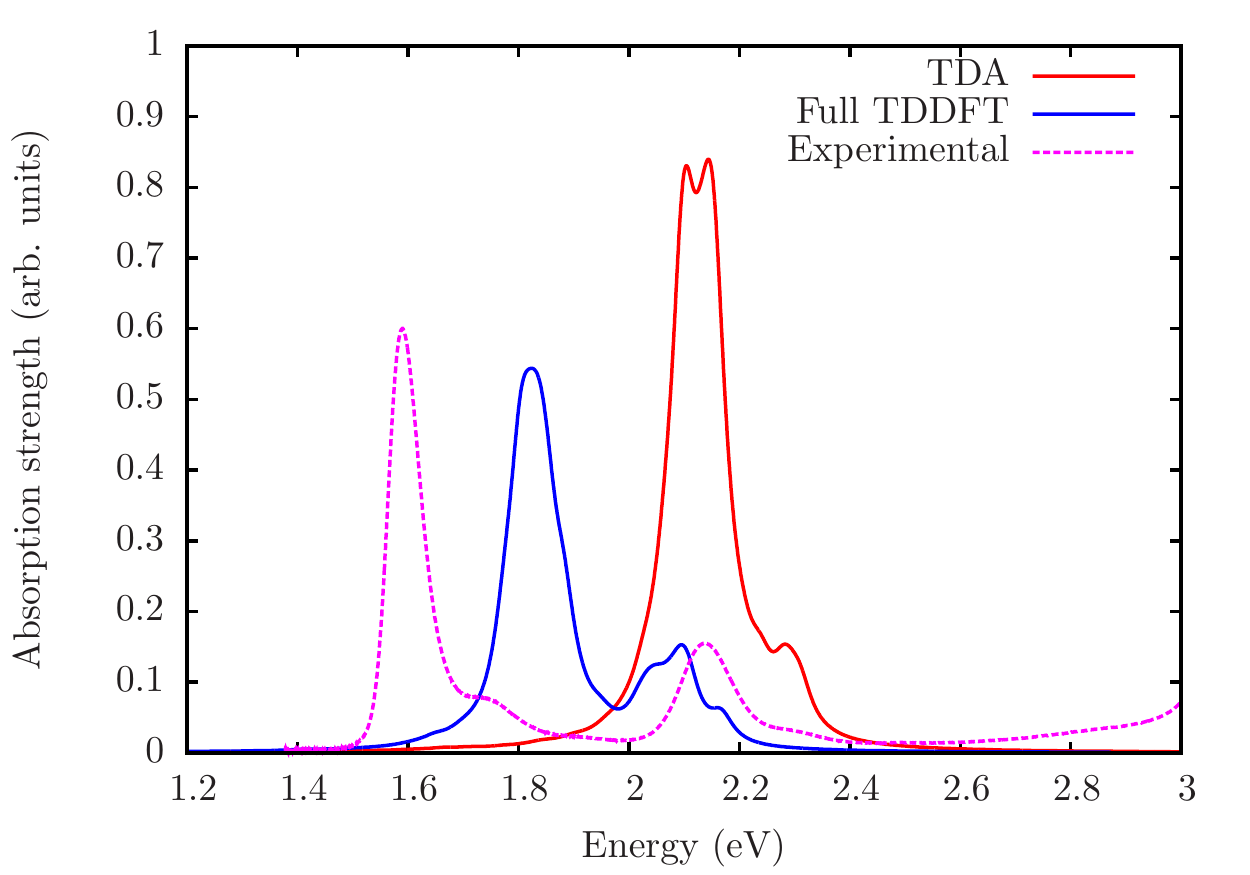}}}
\\
\subfloat[Explicit vs. implicit, averaged snapshots\label{fig:bchl_comparison}]{\resizebox{0.48\textwidth}{!}{\includegraphics{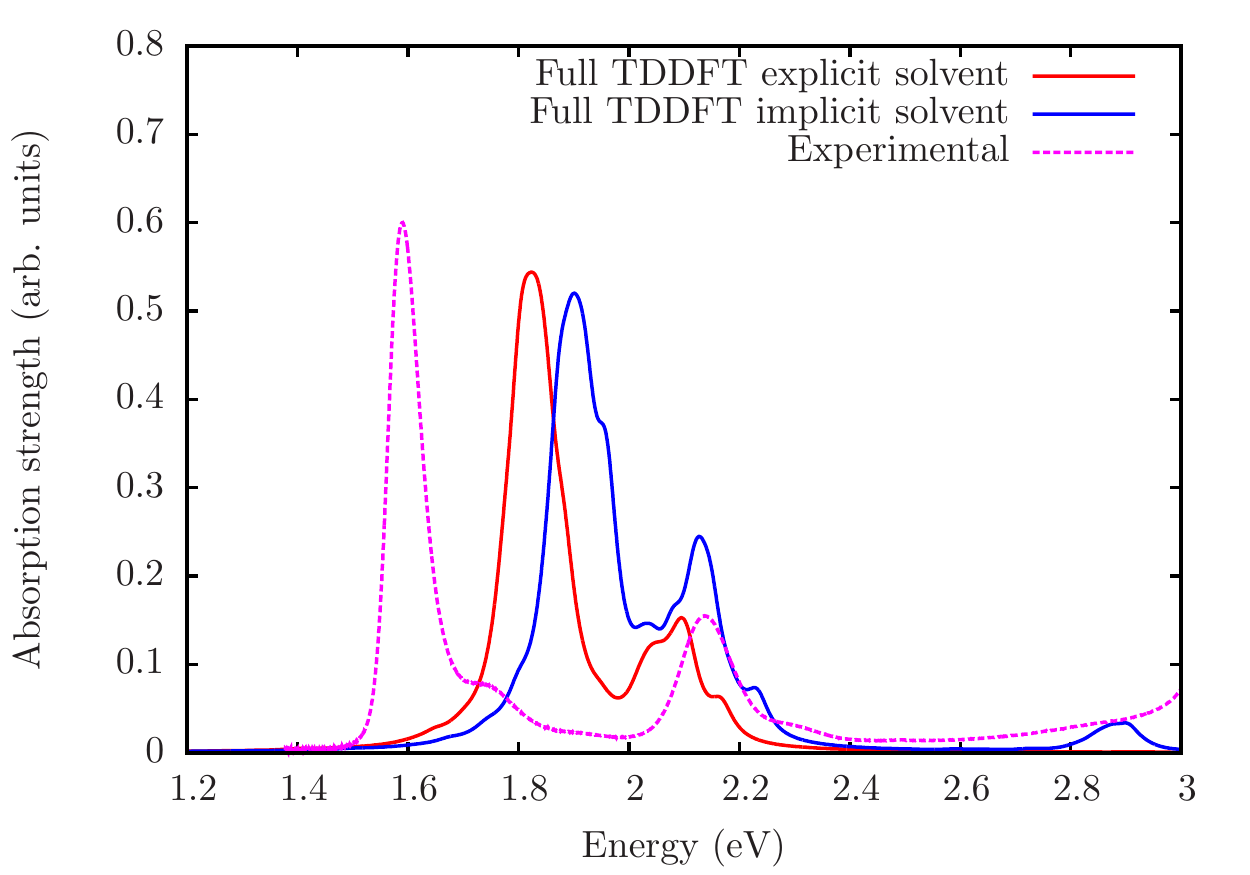}}}
\caption{Absorption spectra of bacteriochlorophyll $a$ as calculated with the TDA and full TDDFT, both using the optimised vacuum structure and the 8 MD snapshots in toluene. Fig. \ref{fig:bchl_implicit} shows the spectrum using the vacuum structure and an implicit solvent model only, while Fig. \ref{fig:bchl_explicit} shows the averaged spectrum of 8 MD snapshots in 15 \AA\, of explicit toluene. Fig. \ref{fig:bchl_comparison} shows a comparison between the averaged spectrum in 15 \AA\, of explicit toluene and the averaged spectrum where the toluene is replaced by an implicit solvent. A Lorentzian broadening of 0.025~eV is used for the TDDFT results and the experimental data is scaled such that the peak height of the Qy transition agrees with that of the Qy peak obtained from full TDDFT in Fig. \ref{fig:bchl_implicit}.}
\label{fig:bchl_combined}
\end{figure}

The results of the TDDFT calculation as compared with the experimental results\footnote{The raw data of the experimental results is taken from http://omlc.org/spectra/PhotochemCAD/html/135.html} can be found in Fig. \ref{fig:bchl_implicit}. As can be seen the experimental spectrum shows three main features: A main absorption peak at around 1.6~eV, a shoulder between 1.65 and 1.75~eV and a second peak at around 2.1~eV. The first and second peaks are commonly referred to as the Qy and Qx transitions respectively and can be characterised as HOMO$\rightarrow$LUMO and HOMO-1$\rightarrow$LUMO transitions in a single-particle picture. As can be seen from Fig.~\ref{fig:bchl_implicit}, the TDA results generate only a single main peak at 2.12~eV that is of Qy character. This main peak shows a shoulder  at 2.05~eV that is mainly of HOMO-1$\rightarrow$LUMO and HOMO-2$\rightarrow$LUMO character. The full TDDFT spectrum on the other hand produces a Qy peak at 1.80~eV that is the lowest excitation of the system, as well as a second peak of Qx character at 1.98~eV, but fails to reproduce a shoulder to the Qy peak. It also shows a third peak with small oscillator strength at 2.11~eV that is of HOMO-2$\rightarrow$LUMO character. 

It can therefore be concluded that the TDA fails in correctly reproducing the absorption spectrum of BChl $a$ in toluene. Not only is the Qy transition overestimated by 0.47~eV compared to the experimental results, it does not correspond to the lowest excitation of the system and there is no clean Qx transition. The full TDDFT results show a considerable improvement. While the Qy transition is still overestimated by 0.2~eV, it now corresponds to the lowest excitation of the system and there is a considerable splitting between the Qy and Qx transitions. However, the full TDDFT results underestimate the energy of the Qx transition by around 0.14~eV compared to experimental results and fail to exhibit a shoulder to the Qy transition. 

The origin of some of the failures of the TDA can be traced by breaking down the excitations into individual Kohn-Sham transitions. In full TDDFT, the Qy transition is almost exclusively (to 95\%, as compared to 60\% in the TDA) a transition between the HOMO and the LUMO. In Bacteriochlorophyll, this transition has a strong dipole moment associated with it, which in turn causes $\textbf{V}_{\textrm{\scriptsize{SCF}}}^{\{1\}}$ to be large and the TDDFT energy to show a large increase compared to the HOMO-LUMO energy difference. In full TDDFT, this large dipole is screened by the de-excitation vector $\textbf{Y}$ which is almost entirely made up of the same HOMO-LUMO transition, thus significantly lowering both the excitation energy and the oscillator strength. In the TDA, $\textbf{Y}=\textbf{0}$ and instead the large dipole moment of the HOMO-LUMO transition is screened by mixing in smaller fractions of higher energy transitions, including fractions of Qx transition. This causes the Qy transition to have a significantly higher energy and larger oscillator strength in the TDA and also contributes to the absence of a clean Qx transition. 

Since optimising the atomic positions of BChl $a$ in vacuum might have lead to a structure that is unrealistic for the system solvated in toluene, it is not clear how much of the failure of full TDDFT to reproduce the experimental results is due to the choice of exchange-correlation functional used in this work. In order to obtain a more realistic structure of BChl $a$ in toluene, we make use of the classical molecular dynamics package AMBER\cite{amber}. We solvate BChl $a$ in 704 molecules of toluene (corresponding to 10,700 atoms for the total system) and equilibrate the system to 300~K and a pressure of 1~atm, followed by a 300~ps simulation in the NVE ensemble. From this MD run we take 8 snapshots 10~ps apart that are then used as input atomic positions in the TDDFT calculations. In order to include solvent effects at a quantum mechanical level, we explicitly include all toluene molecules within a 15~\AA \, radius from the Mg atom in the calculation, while representing the rest of the solvent by an implicit solvation model. This process yields a system size of 700-800 atoms depending on the snapshot, which is closer to the limit of sizes that can be treated by conventional $\mathcal{O}(N^3)$ methods. The structure of the BChl $a$ molecule as obtained from a single MD snapshot can be found in Fig.~\ref{fig:toluene_struct}. As can be seen, the alkane tail extends away from the porphine ring in this configuration, while the ring itself is no longer perfectly flat. 

The averaged full TDDFT and TDA spectra of the solvated MD snapshots as compared with the same experimental dataset are shown in Fig.~\ref{fig:bchl_explicit}. Note that the full TDDFT results now show the Qy transition at 1.83~eV and the Qx transition at 2.10~eV. While the Qy transition is still overestimated by 0.2~eV, the shape of the feature is considerably improved, as the Qx transition is now in very good agreement with experimental results, both in its positioning and in its intensity. It is worth pointing out that a number of snapshots also show a shoulder to the main Qy transition that is of HOMO-2$\rightarrow$LUMO character. However, in the averaged spectrum this feature is not as pronounced as in the experimental spectrum, which can be due to the fact that the splitting between the Qx and Qy transitions is too small compared to experiment. The TDA results on the other hand still fail to reproduce main spectral features. The Qy transition is overestimated by 0.5~eV, although the shoulder present in Fig. \ref{fig:bchl_implicit} has disappeared. The spectrum shows a new peak at approximately 2.3~eV that does however have a different character to the Qx transition in full TDDFT. A clearly identifiable Qx transition is still absent from the TDA results. It can therefore be summarised that full TDDFT yields a much improved representation of the experimental results at the PBE level as long as a realistic structure for the solute and the solvent environment is obtained. The main failure of full TDDFT for this system is the overestimation of the Qy transition by 0.2~eV which can most likely be ascribed to errors in the exchange-correlation functional used. 

While the 700-800 atom systems obtained from classical MD yield a relatively good spectral shape for full TDDFT at the PBE level, they are considerably larger than the 140 atoms of the solute alone. It is therefore worth investigating how much of the improvement of the spectrum from Fig.~\ref{fig:bchl_implicit} to Fig.~\ref{fig:bchl_explicit} is due to the different ionic positions of the solute and how much is due to an explicit quantum mechanical treatment of solvent molecules in the calculation. For this purpose we take the atomic positions of BChl $a$ from the 8 MD snapshots and compute the absorption spectrum in implicit solvent, without including any explicit representation of toluene molecules. The result can be found in Fig.~\ref{fig:bchl_comparison}. As can be seen, the positioning of the Qx transition in implicit solvent is in very good agreement with the experimental data. However, its oscillator strength is significantly overestimated. Furthermore, the explicit solvent environment causes the Qy transition to red-shift by about 0.1~eV. It can be concluded that while most of the improvements in spectral features compared to Fig.~\ref{fig:bchl_implicit} are due to the more realistic atomic positions of the Bacteriochlorophyll in toluene, the explicit inclusion of the toluene environment at the TDDFT level yields a spectrum that is in closest agreement with experimental results. 

In conclusion it can be summarised that full TDDFT at the PBE level reproduces experimental results to a satisfactory degree, while the TDA completely fails in this system. The best agreement between experiment and calculation is obtained when taking the atomic positions of an MD snapshot of BChl in toluene and including the local solvent environment explicitly in the TDDFT calculation. While an explicit treatment of the solvent molecules requires large scale TDDFT calculations, it is demonstrated that the method presented in this work is well suited for tackling these systems, opening up the possibility of more detailed studies of solvent effects on chromophores. 

\subsection{Linear-scaling capabilities}
\label{sec:linear_scaling}

\begin{figure}
\centering
\resizebox{0.48\textwidth}{!}{\includegraphics{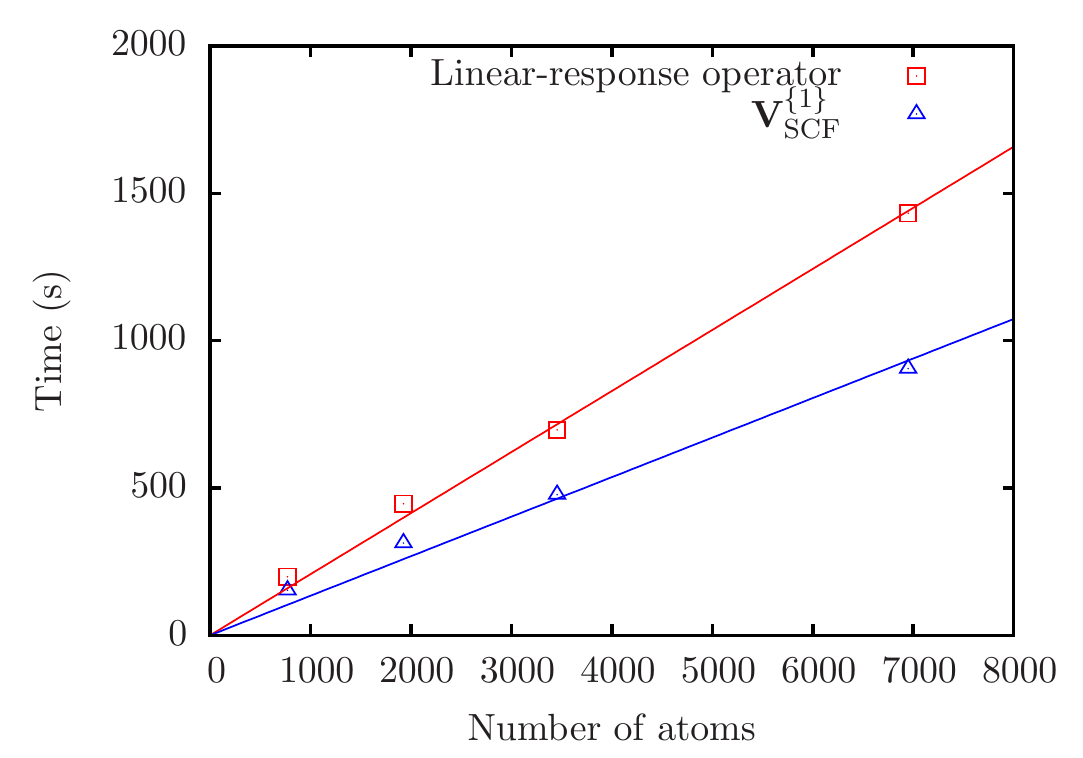}}
\caption{Time taken for applying the TDDFT operator and calculating $\textbf{V}_{\textrm{\scriptsize{SCF}}}^{\{1\}}$ for different system sizes of bacteriochlorophyll $a$ in toluene. The lines shown are linear fits. Calculations were performed on eight SandyBridge nodes containing 16 cores each.}
\label{fig:timing}
\end{figure}

We now focus on demonstrating the linear-scaling capabilities of the TDDFT method presented in this work. For this purpose, we take one of the classical MD snapshots studied in the previous section and compute the Qy transition of BChl $a$, including progressively larger regions of the solvent environment. Calculations are performed using the same parameters as in the previous section, apart from using a smaller radius of 10~$a_0$ for the NGWF set $\{\chi_\alpha\}$. In order to reach linear-scaling computational effort with system size, it is necessary to truncate all involved density matrices $\textbf{P}^{\{v\}}$, $\textbf{P}^{\{c\}}$, $\textbf{P}^{\{p\}}$ and $\textbf{P}^{\{q\}}$ to make matrix-matrix operations scale as $\mathcal{O}(N)$. Here, we choose a tight truncation radius of 20~$a_0$, causing the unoccupied and ground-state density matrices to have the same sparsity pattern as the NGWF representation overlap matrices $\textbf{S}^\chi$ and $\textbf{S}^\phi$. However, by performing a full TDDFT calculation on the 770-atom system of the previous section, we have confirmed that the error introduced to the energy of the Qy transition for such a cutoff is only 0.03~eV. 

While the Qy transition of interest in this system retains a relatively localised character and the error introduced through truncating $\textbf{P}^{\{p\}}$ and $\textbf{P}^{\{q\}}$ can be expected to be relatively small, the total error of 0.03~eV is a combination of errors introduced by the response density matrices and by the truncation of the ground state density matrix. The calculations show, that for relatively localised excitations, fully linear-scaling calculations on realistic systems are possible with only introducing minor errors. For very delocalised excitations, the truncation of the response density matrix becomes more difficult and the effects of applying a truncation on long-range charge-transfer excitations has been discussed elsewhere in more detail\cite{tddft_onetep}. Here we note, that for a large class of systems like pigment-protein complexes, excitations of interest are expected to retain a relatively localised character and fully linear-scaling calculations are indeed possible.

\begin{figure}
\centering
\resizebox{0.48\textwidth}{!}{\includegraphics[natwidth=5in,natheight=3.5in]{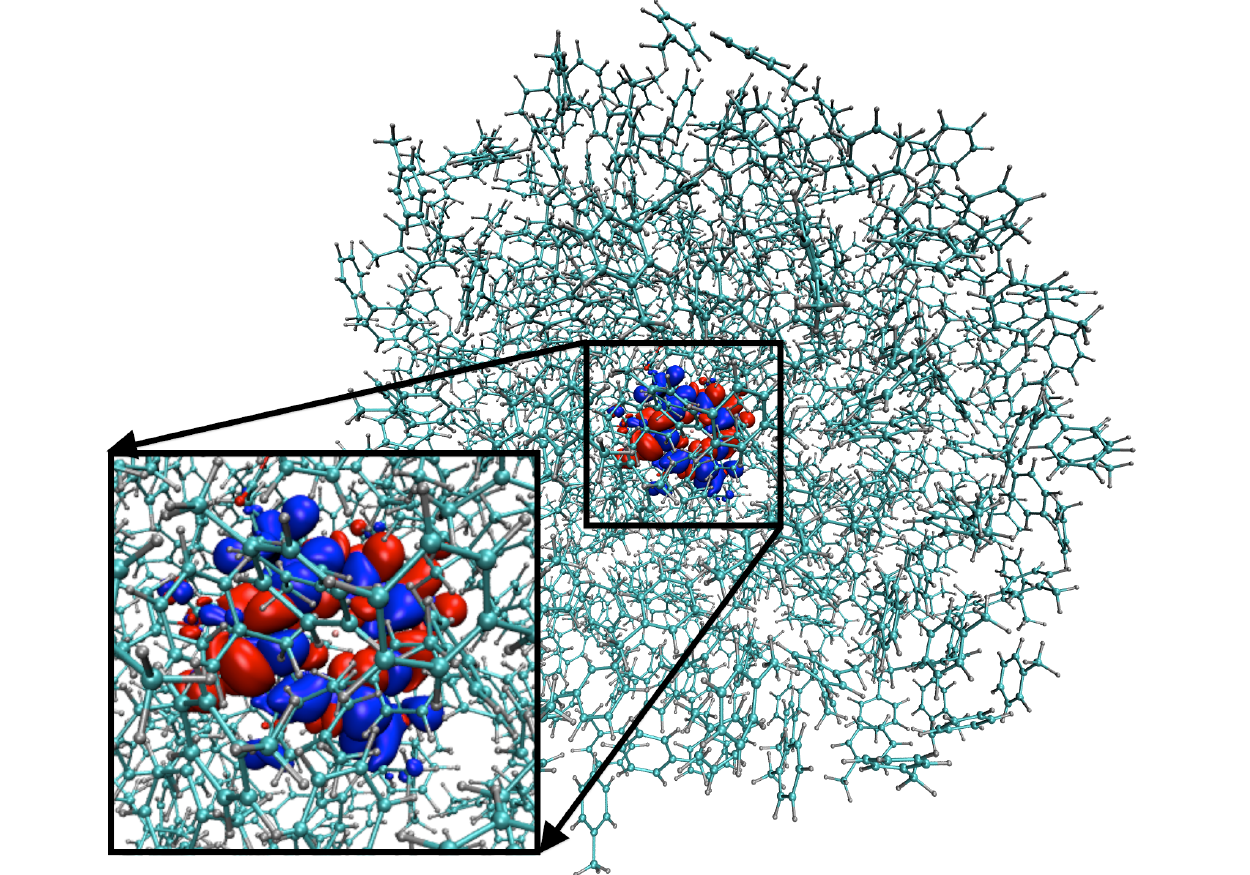}}
\caption{The electron-hole difference density of the Qy transition of bacteriochlorophyll surrounded by a sphere of 30~\AA \, radius of toluene solvent molecules. The calculation includes 454 toluene molecules, corresponding to a total system size of 6950 atoms. The figure was created using VMD\cite{vmd}.}
\label{fig:qy_trans}
\end{figure}

We choose to perform a linear-scaling test of the full TDDFT method on four different system sizes of Bchl in Toluene, each specified by the radius, as measured from the Mg atom, up to which solvent molecules are included in the calculation. The four different radii chosen are 15, 20, 24 and 30~\AA, corresponding to system sizes of 770, 1925, 3455 and 6950 atoms. The timings, both for evaluating the self-consistent field response $\textbf{V}_{\textrm{\scriptsize{SCF}}}^{\{1\}}$ and applying the full TDDFT operator (\ref{eqn:operator_RPA}) are shown in Fig.~\ref{fig:timing}. In general, a clear linear trend can be observed for both the calculation of $\textbf{V}_{\textrm{\scriptsize{SCF}}}^{\{1\}}$ and the TDDFT operator, with slight discrepancies for the two smaller systems. These discrepancies are likely originating from the fact that matrices in the two smallest systems are still relatively dense and a full transition to linear scaling effort only occurs at larger system sizes. However, it becomes clear that the algorithm is capable of solving the full TDDFT eigenvalue equation for systems of thousands of atoms in linear-scaling effort. Furthermore, recent work implementing hybrid OpenMP-MPI approaches to parallelism mean that these calculations scale efficiently to many thousands of CPU cores\cite{omp_mpi}.

Figure~\ref{fig:qy_trans} shows a plot of the electron-hole difference density for the fully converged Qy transition of the 6950 atom system. It should be noted that these system sizes are inaccessible by standard $\mathcal{O}(N^3)$ approaches, both at the DFT and the TDDFT level. The calculation presented here is to be seen for demonstration purposes only, given that the Qy transition retains a relatively localised character and a fully quantum mechanical treatment of such a large region of the solvent environment can be considered unnecessary. However, it should be pointed out that the system treated here is of similar size as a variety of pigment-protein complexes, most notably the Fenna-Matthews-Olson complex, where seven BChl molecules are embedded in a complex protein environment and site energy variations due to environmental screening effects are both subtle and important\cite{shim_fmo, neugebauer_fmo, danny_fmo}. The method presented in this work has the potential of treating these systems fully quantum mechanically, without relying on semi-classical approximations. 

\section{Conclusion}
In conclusion, we have outlined a preconditioned conjugate gradient algorithm capable of solving for the lowest eigenvalues of the full TDDFT equation with linear-scaling effort. We have demonstrated the efficiency of the compact NGWF representation and shown that it yields results comparable to those obtained with well-converged Gaussian basis sets. Furthermore, the vital importance of preconditioning the iterative solution to the TDDFT equation has been demonstrated, yielding a four-fold speedup of the convergence rate in the case of \emph{trans}-azobenzene. 

We have furthermore shown that the TDA fails to reproduce experimental results of BChl $a$ in toluene solution at the PBE level, while full TDDFT yields a significant improvement of all spectral features. It was also shown that the best results compared to experimental data are achieved when treating a certain part of the solvent explicitly at TDDFT level, making it necessary to perform TDDFT calculations of several hundreds of atoms. While these system sizes can reach the the limits of standard $\mathcal{O}(N^3)$ methods, they are straightforwardly treated with the method introduced here, opening up the possibility of more detailed studies on the effects of pigment-solvent interactions on excitation energies of chromophores. 

Finally, we have shown that the algorithm scales fully linearly with system size as long as all involved density matrices are truncated. The largest full TDDFT calculation performed in this work treats a system of 6950 atoms, far larger than systems that can be realistically addressed with cubic-scaling approaches. These large-scale systems are of the same order of magnitude as a large variety of pigment-protein complexes that are studied in the field of computational biology, opening up the possibility of computing their excitation spectra without the need to rely on any semi-classical approximations. 

\begin{acknowledgements}
TJZ acknowledges the support of EPSRC Grant EP/J017639/1 and the ARCHER eCSE programme. MCP and PDH acknowledge the support of EPSRC grant EP/J015059/1. The underlying data of this publication can be accessed via the following persistent URI: https://www.repository.cam.ac.uk/handle/1810/251293
\end{acknowledgements}

\end{document}